\documentclass[longauth]{aa} 

\usepackage{graphics,graphicx,graphbox}
\usepackage{rotating}
\usepackage[T1]{fontenc}

\usepackage{natbib}
\usepackage{ulem}
\usepackage[flushleft]{threeparttable}
\usepackage{multirow}
\usepackage{gensymb}
\usepackage{scrextend}   
\usepackage[colorlinks=true, linkcolor=blue, citecolor=blue, draft=false]{hyperref}
\usepackage{xcolor}
\usepackage{orcidlink}
\usepackage{caption}
\usepackage{subcaption}
\usepackage{float}  
 
\newcommand{\mstar}{$M_\star$}
\newcommand{\mgas}{$M_{\rm gas}$}
\newcommand{\fgas}{$\mu_{\rm gas}$}
\newcommand{\taugas}{$\tau_{\rm gas}$}
\newcommand{\msunyr}{M$_\odot$\,yr$^{-1}$}
\newcommand{\msun}{M$_\odot$}
\newcommand{\lir}{$L_{\rm IR}$}

\newcommand{\Kkmspc}{K\,km\,s$^{-1}$\,pc$^2$}
\newcommand{\rsb}{$R_{\rm SB}$}
\newcommand{\acosmos}{A$^3$COSMOS}

\newcommand{\fgasMed}{1.32}
\newcommand{\taugasMed}{0.78} 

\def\zhou#1 {{\textcolor{red}{#1}}\ }

\begin{document} 
      \title{Noema formIng Cluster survEy (NICE): \\ A census of star formation and cold gas properties in massive protoclusters at 1.5\,<\,$z$\,<\,4 }
   \titlerunning{NICE-COSMOS: Star Formation and Cold Gas Properties}

\author{Luwenjia Zhou\inst{\ref{c1},\ref{c2}}\orcidlink{0000-0003-1687-9665}
\and Tao Wang\inst{\ref{c1},\ref{c2}}\thanks{E-mail: \texttt{taowang@nju.edu.cn}}
\and Emanuele Daddi\inst{\ref{c3}} 
\and Rosemary Coogan\inst{\ref{c3}} 
\and Hanwen Sun \inst{\ref{c1},\ref{c2}} 
\and Ke Xu \inst{\ref{c1},\ref{c2}} 
\and Vinodiran Arumugam \inst{\ref{c8}}
\and Shuowen Jin\inst{\ref{c4},\ref{c5}}
\and Daizhong Liu\inst{\ref{c7}} 
\and Shiying Lu\inst{\ref{c-14},\ref{c-15},\ref{c2} }
\and Nikolaj Sillassen \inst{\ref{c4},\ref{c5}}
\and Sicen Guo \inst{\ref{c3}} 
\and Guillaume Elias \inst{\ref{c3}} 
\and Yijun Wang\inst{\ref{c1},\ref{c2}} 
\and Yong Shi\inst{\ref{c1},\ref{c2}}
\and Zhi-Yu Zhang\inst{\ref{c1},\ref{c2}}
\and Qinghua Tan\inst{\ref{c7}}
\and Qiusheng Gu\inst{\ref{c1},\ref{c2}}
\and David Elbaz \inst{\ref{c3}}
\and Aurelien Henry \inst{\ref{c3}}
\and Benjamin Magnelli \inst{\ref{c3}}
\and Carlos G{\'o}mez-Guijarro \inst{\ref{c3}}
\and Chiara d'Eugenio \inst{\ref{c-2}, \ref{c-3}}
\and Georgios E. Magdis \inst{\ref{c4},\ref{c5},\ref{c-1}}\orcidlink{0000-0002-4872-2294}
\and Francesco Valentino \inst{\ref{c-4}}\orcidlink{0000-0001-6477-4011}
\and Zhiyuan Ji \inst{\ref{c-5}}\orcidlink{0000-0001-7673-2257}
\and Raphael Gobat \inst{\ref{c-6}}
\and Ivan Delvecchio \inst{\ref{c-7}}\orcidlink{0000-0001-8706-2252}
\and Mengyuan Xiao \inst{\ref{c-8}}\orcidlink{0000-0003-1207-5344}
\and Veronica Strazzullo \inst{\ref{c-9},\ref{c-99}}
\and Alexis Finoguenov \inst{\ref{c-10}}
\and Eva Schinnerer \inst{\ref{c6}}
\and R. Michael Rich \inst{\ref{c-11}}
\and Jiasheng Huang \inst{\ref{c-13}}
\and Yu Dai \inst{\ref{c-13}}
\and Yanmei Chen\inst{\ref{c1},\ref{c2}}
\and Fangyou Gao\inst{\ref{c1},\ref{c2}}
\and Tiancheng Yang\inst{\ref{c1},\ref{c2}}
\and Qiaoyang Hao\inst{\ref{c1},\ref{c2}}
}

\institute{School of Astronomy and Space Science, Nanjing University, Nanjing 210093, China \label{c1}
\and Key Laboratory of Modern Astronomy and Astrophysics (Nanjing University), Ministry of Education, Nanjing 210093, China \label{c2}
\and AIM, CEA, CNRS, Universit\'{e} Paris-Saclay, Universit\'{e} Paris Diderot, Sorbonne Paris Cit\'{e}, F-91191 Gif-sur-Yvette, France \label{c3}
\and IRAM, 300 rue de la piscine, F-38406 Saint-Martin d'H\`{e}res, France\label{c8}
\and Cosmic Dawn Center (DAWN), Jagtvej 128, DK2200 Copenhagen N, Denmark \label{c4}
\and DTU-Space, Technical University of Denmark, Elektrovej 327, 2800 Kgs. Lyngby, Denmark \label{c5}
\and Max-Planck-Institut f\"{u}r Extraterrestrische Physik (MPE), Giessenbachstrasse 1, 85748 Garching, Germany \label{c6}
\and School of Mathematics and Physics, Anqing Normal University, Anqing 246133, China \label{c-14}
\and Institute of Astronomy and Astrophysics, Anqing Normal University, Anqing 246133, China \label{c-15}
\and Purple Mountain Observatory, Chinese Academy of Sciences, 10 Yuanhua Road, Nanjing 210023, China \label{c7}
\and Niels Bohr Institute, University of Copenhagen, Jagtvej 128, DK-2200 Copenhagen N, Denmark\label{c-1}
\and Instituto de Astrofísica de Canarias, C. V\'{i}a L\'{a}ctea, s/n, 38205 La Laguna, Tenerife, Spain\label{c-2}
\and Universidad de La Laguna, Dpto. Astrofísica, 38206 La Laguna, Tenerife, Spain\label{c-3}
\and European Southern Observatory, Karl-Schwarzschild-Str. 2, D-85748 Garching bei Munchen, Germany\label{c-4}
\and Steward Observatory, University of Arizona, 933 N. Cherry Avenue, Tucson, AZ 85721, USA\label{c-5}
\and INAF – Osservatorio di Astrofisica e Scienza dello Spazio di Bologna, Via Gobetti 93/3, I-40129 Bologna, Italy \label{c-7}
\and Department of Astronomy, University of Geneva, Chemin Pegasi 51, 1290 Versoix, Switzerland\label{c-8}
\and Instituto de F\'{i}sica, Pontificia Universidad Cat\'{o}lica de Valpara\'{i}so, Casilla 4059, Valpara\'{i}so, Chile\label{c-6}
\and INAF - Osservatorio Astronomico di Trieste, Via Tiepolo 11, 34131, Trieste, Italy \label{c-9}
\and IFPU - Institute for Fundamental Physics of the Universe, Via Beirut 2, 34014, Trieste \label{c-99}
\and  Department of Physics, University of Helsinki, Gustaf H\"{a}llstr\"{o}min katu 2, FI-00014 Helsinki, Finland\label{c-10}
\and Department of Physics \& Astronomy, University of California Los Angeles, 430 Portola Plaza, Los Angeles, CA 90095, USA\label{c-11}
\and Chinese Academy of Sciences South America Center for Astronomy (CASSACA), National Astronomical Observatories of China (NAOC), 20A Datun Road \label{c-13}
}
\date{Received --; accepted --}

 
  \abstract{
Massive protoclusters at $z \sim 1.5-4$, the peak of the cosmic star formation history, are key to understanding the formation mechanisms of massive galaxies in today's clusters. However, studies of protoclusters at these high redshifts remain limited, primarily due to small sample sizes and heterogeneous selection criteria. For this work, we conducted a systematic investigation of the star formation and cold gas properties of member galaxies of eight massive protoclusters in the COSMOS field , using the statistical and homogeneously selected sample from the Noema formIng Cluster survEy (NICE). Our analysis reveals a steep increase in the star formation rates per halo mass ($\Sigma_{\rm SFR} /M_{\rm halo}$) with redshifts  in these intensively star-forming protoclusters, reaching values one to two orders of magnitude higher than those observed in the field at $z$\,>\,2. We further show that instead of an enhancement of starbursts, this increase is largely driven by the concentration of massive and gas-rich star-forming galaxies in the protocluster cores. The member galaxies still generally follow the same star-forming main sequence as in the field, with a moderate enhancement at the low-mass end. Notably, the most massive protocluster galaxies (\mstar\,>\,8$\times$10$^{10}$\msun) exhibit higher \fgas\, and \taugas\,   than their field counterparts, while remaining on the star-forming main sequence.  These gas-rich, massive, and star-forming galaxies are predominantly concentrated in the protocluster cores and are likely progenitors of massive ellipticals in the center of today's clusters. These results suggest that the formation of massive galaxies in such environments is sustained by substantial gas reservoirs, which in turn support persistent star formation and drive early mass assembly in forming cluster cores.
 }


   \maketitle
%
\section{Introduction}
The environmental influence on the progenitors of massive galaxies in local clusters is critical for understanding the formation of these galaxies  \citep{Peng2010, Behroozi2019, vanderBurg2020} . Over the past decade, extensive studies of massive protoclusters at  the peak of the cosmic star formation history ($z \sim 2-4$), selected through various methods, have been inconclusive in producing an unambiguous consistent picture of the environmental effects on star formation and cold gas properties of galaxies in these structures \citep[e.g.,][]{Rudnick2017, Gomez-Guijarro2019, Jin2023, Jin2024, Bakx2024, Pensabene2024}.
Specifically, some structures show a prevalence of submillimeter galaxies (SMGs) or ultra-luminous infrared galaxies (LIRGs), suggesting enhanced star formation in their galaxies as compared to average  field environments \citep[e.g.,][]{Casey2016, Wang2016, Oteo2018}.
Other studies, primarily based on H$\alpha$ emitters in protoclusters, report little to no difference in the star-forming main sequence (SFMS)  compared to their field counterparts  \citep[e.g,][]{Valentino2015, Shimakawa2018,  Perez-Martinez2023}.
Molecular gas is the fuel for star formation. The well-established star formation law \citep{Kennicutt1998, Gao2004} highlights the role of internal secular processes in the conversion of cold gas into stars. Additionally, a bimodal scenario has been proposed to describe the observed offset between starburst galaxies and normal star-forming galaxies, where starburst galaxies show higher star formation efficiency (SFE), which may result from a higher merger rate or a more compact distribution of molecular gas, among other factors \citep{Daddi2010b, Sargent2014, Liu2019}.

In dense environments, increased interactions can compress gas, thus enhancing SFEs, while cold gas inflows through filaments connected to protoclusters can replenish gas reservoirs, thus sustaining high gas fractions.   Deep CO observations of cluster galaxies $z$\,$\sim$\,1.6 reveal a consistent or enhanced gas fraction compared to the field counterparts \citep{Rudnick2017, Noble2017}. 
In the Spiderweb protocluster at $z$=2.15, \citet{Dannerbauer2017} identified a massive and extended CO(1-0) disk around the central radio galaxies. Follow-up analysis by \citet{Perez-Martinez2025} revealed a decreasing gas fraction with increasing stellar mass, showing a particularly steep decline above log(\mstar/\msun)\,=\,10.5. 
In CL~J1001, a starbusting protocluster at $z$\,=\,2.51, \citet{Wang2018} observed  decreasing gas fractions toward the cluster center, suggesting the imminent formation of a passive core by $z\sim$\,2.  \citet{Tadaki2019} 
 found a declining gas fraction and gas depletion time with increasing stellar mass in galaxies in three protoclusters traced by radio galaxies at $z\sim 2$, and suggested that gas accretion is accelerated in less massive galaxies and suppressed in more massive ones.
At even higher redshifts ($z\gtrsim$\,4),  extreme starbusting protocluster cores such as SPT2349-56 and Distant Red Core (DRC) exhibit lower gas fractions compared to field SMGs \citep{Long2020, Hill2022}.
Nevertheless, there is a paucity of protoclusters at high redshift with deep multi-wavelength coverage to comprehensively link the stellar and gas properties of member galaxies. Such observations are essential for revealing the physical mechanisms driving the formation of massive galaxies in these dense environments. 
 
The Noema formIng Cluster survEy (NICE, \citealt{Zhou2024}) is a 159-hour NOEMA Large Program (ID: M21AA, PIs: E. Daddi and T. Wang) targeting 48 massive protocluster candidates across multiple fields, complemented by a 40-hour ALMA program (ID: 2021.1.00815.S) focusing on 25 southern candidates. The primary goal of these programs is to spectroscopically confirm protoclusters at cosmic noon through CO emission from their member galaxies, significantly expanding the sample size in this field of research.
Notably, eight protoclusters at 1.5\,<\,$z$\,<\,4 have been confirmed in the COSMOS field \citep{Sillassen2024}. The COSMOS field provides extensive ancillary data across optical, near-infrared (NIR), far-infrared (FIR), submillimeter, and radio wavelengths \citep{Weaver2022, Jin2018, Smolcic2017}, which, combined with the CO observations from the NICE survey,  offer a unique opportunity to investigate the environmental impact on galaxy evolution.

For this paper we investigated the star formation activity and cold gas content of eight massive protoclusters in the COSMOS field, identified through the NICE survey (hereafter NICE-COSMOS).  Our analysis focuses on the protoclusters' central region (within one virial radius), as detailed in the following sections.
We adopt a standard $\Lambda$CDM cosmology with $H_0$\,=\,70\,km\,s$^{-1}$\,Mpc$^{-1}$, $\Omega_{\rm m}$\,=\,0.3, and $\Omega_{\rm \Lambda}$\,=\,0.7, and a \citet{Chabrier2003} initial mass function. 
 
\section{Data}
\label{sec:data}
\subsection{NICE-COSMOS protoclusters}
The eight protoclusters in NICE-COSMOS, along with their member identification, are detailed in \citet{Sillassen2024}. Briefly,  the members are selected with |$z_{\rm phot}$ -  $z_{\rm spec,group}$|\,<\,0.1(1\,+\,$z_{\rm spec,group}$) within a virial projected radius  based on the COSMOS2020 catalog \citep{Weaver2022}.  
The NICE survey selects intensively star-forming massive protoclusters at 2\,$\lesssim$\,$z$\,$\lesssim$\,4, which are chosen as overdensities of high-$z$ massive galaxies traced by red IRAC sources in association with SPIRE-350$\mu$m peakers that show intense star formation. We list the selection criteria here:

(1) Overdensity of red IRAC sources:
\begin{equation}
\begin{split}
[3.6] - [4.5] & > 0.1 \,; \\
20 < [4.5] & < 23 \,; \\
\Sigma_N & > 5\sigma \, \text{(N = 5 or 10)}.
\end{split}
\label{eq:irac}
\end{equation}
[3.6], [4.5] are IRAC magnitudes from COSMOS2020 \citep{Weaver2022}, and $\Sigma_{N}$\,=\,$\frac{N}{\pi\,d_N^2}$ is the Nth closest neighbor density estimator,  where $d_N$ is the distance to the Nth closest galaxy. 

(2) SPIRE-350$\mu$m peakers:
\begin{equation}
\begin{split}
S_{\rm 500} & > 30\,\rm mJy \,; \\
S_{\rm 350}/S_{\rm 250} & > 1.06 \,; \\
S_{\rm 500}/S_{\rm 350} & > 0.72 \,.
\end{split}
\label{eq:spire}
\end{equation}
$S_{\rm 250}$, $S_{\rm 350}$, and $S_{\rm 500}$ are the Herschel/\textit{SPIRE} fluxes at 250$\mu$m, 350$\mu$m, and 500$\mu$m.

Figure~\ref{fig:img-rgb}  shows the spatial distribution of all members including CO and continuum detections in the eight protoclusters in NICE-COSMOS, which are primarily concentrated at the cluster centers.  For the following analysis, we focus on galaxies within one virial radius, 1$R_{\rm vir}$, as defined in \citep{Sillassen2024}, as this area is also well covered by the NOEMA and ALMA observations. The virial radii range from 141\,pkpc to 429\,pkpc, and the number of member galaxies per structure varies between 13 and 31.

According to the selection method, the NICE protoclusters are traced by intensively star-forming massive galaxies, which are likely to undergo a more active formation phase driven by enhanced gas reservoirs, elevated merger rates, or other physical processes, compared to protoclusters selected through alternative methods.

\subsection{Stellar masses and star formation rates}
\label{sec:mstar_sfr}
We used stellar masses and star formation rates (SFRs) derived with \texttt{LePhare} \citep{Arnouts2002} in COSMOS2020 \citep{Weaver2022}. For galaxies detected in the FIR, specifically those in the super-deblended catalog \citep{Jin2018}, we also explored various SFR tracers to obtain accurate measurements. 

 We first cross-matched the members in the super-deblended catalog \citep{Jin2018} with a separation smaller than 1$\arcsec$. The matches have a median separation of 0.15$\arcsec$. Notably, the 3GHz and 24$\mu$m fluxes are measured directly at the galaxy positions, hence the risk of misidentification or flux misattribution is substantially reduced.   For galaxies detected at 3\,GHz \citep{Smolcic2017} with a signal-to-noise ratio (S/N) greater than 3, we calculated the SFR from \lir \,\citep{Kennicutt1998}, as derived from the 3 GHz radio emission.  This was done using the infrared radio correlation  \citep[IRRC, $q_{\rm IR}(z,M_\star)$,][]{Delvecchio2021} and a spectral index of $\alpha$\,=\,-0.8. 
The high spatial resolution (0.75\arcsec) of the VLA 3\,GHz observations helps mitigate blending issues in crowded regions, which can affect FIR flux measurements. For sources with S/N$_{\rm 3GHz}$\,<\,3, we computed \lir \, from 24$\mu$m fluxes using the conversion from \citet{Wuyts2008}. The results are listed in Table~\ref{tab:properties}. 

\begin{figure}
\centering
\includegraphics[align=c, width=0.9\linewidth]{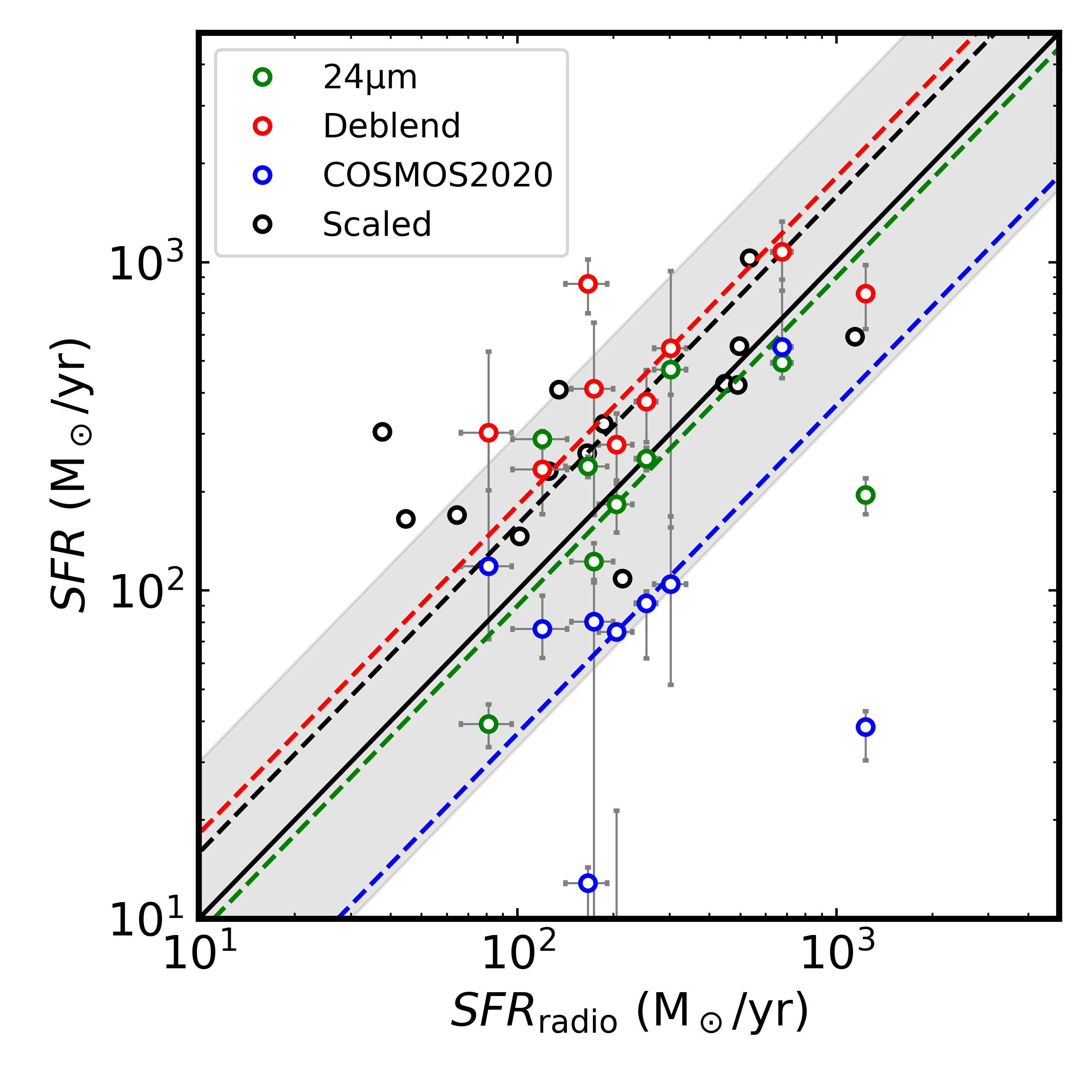}
\caption{Comparison of SFRs derived from radio emission with those derived using other methods.
The comparison includes SFRs estimated from \citet[green]{Wuyts2008} , deblended FIR fluxes from  \citet[red]{Jin2018} , SFRs derived with \textsc{LePhare} as in COSMOS2020 \citep[blue]{Weaver2022},  and  SFRs scaled from overall average SEDs of the protoclusters to dust continuum (black). Only galaxies with S/N$_{\rm 3GHz}$\,>\,3 are shown. The black solid line represents a 1:1 relation, while the shaded region indicates the 0.2\,dex uncertainties. The dashed colored lines show the median ratios between \textit{SFR}$_{\rm radio}$ and the SFRs derived from the fours different methods, respectively: 0.90, 1.81, 0.37, 1.57. 
}  
\label{fig:sfr_compare}
\end{figure}

As we show in Fig.~\ref{fig:sfr_compare}, \textit{SFR}$_{\rm radio}$  and \textit{SFR}$_{\rm 24\mu m}$ are consistent. For galaxy 839791 in COSMOS-SBC6, which has a radio luminosity twice that of the crossover luminosity described in  \citet{Wang2024},    the threshold above which radio emission is dominated by AGN,  we used \textit{SFR}$_{\rm 24\mu m}$ to  minimize the contamination from AGN-related radio emission. 

In addition, we calculated SFRs  by scaling the total \lir\, derived from the best-fit SED of the integrated FIR photometry of each protocluster \citep{Sillassen2024} to the dust continuum emission of individual members, assuming the same dust temperature for all members. We find that the scaled SFR is generally 1.6 times higher than \textit{SFR}$_{\rm radio}$; this is more prominent for those galaxies with $SFR_{\rm radio}\sim$100\msunyr, while galaxies with  higher \textit{SFR}$_{\rm radio}$ have similarly scaled SFRs.  

Overall, we conclude that  using \textit{SFR}$_{\rm radio}$ does not change the conclusion of this paper.

\subsection{Gas masses} 
\label{sec:gasmass}

The molecular gas mass is converted from the Rayleigh-Jeans dust continuum at 850$\mu$m, $M_{\rm gas}^{\rm RJ}$, as calibrated with the prescription  in \citet{Hughes2017}. This method leverages dust continuum emission, which is intrinsically associated with gas in the entire cold molecular gas reservoir \citep{Genzel2015, Scoville2016}, and offers a much more sensitive measure than high-$J$ CO emission, which primarily traces the dense gas phase only \citep{Liu15, Daddi2015, Valentino2020c} and  has limited detections due to the narrow line width compared to continuum observations.
The dust continuum at 3 mm and 2 mm from the NOEMA and ALMA observations of the NICE survey, closely corresponding to rest-frame 850\,$\mu$m for the cluster galaxies, was used to derive gas masses. The NOEMA and ALMA observations have average angular resolutions of 4.1\arcsec and 0.7\arcsec, respectively. For spatially resolved sources, we used a Gaussian model to fit the visibilities. Unresolved sources were modeled as point sources.

\begin{figure}
\centering
\includegraphics[align=c, width=0.9\linewidth]{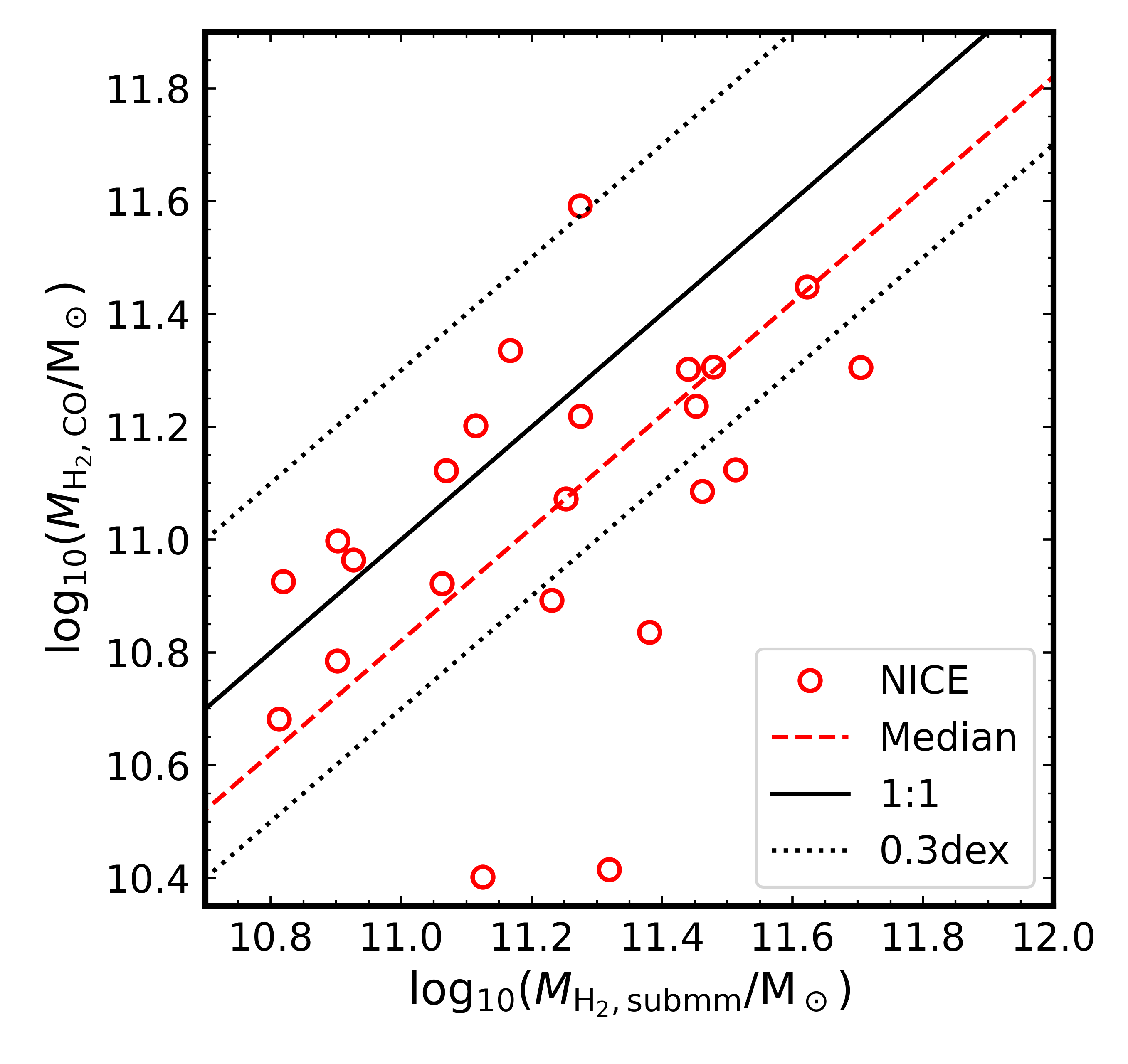}
\caption{Comparison of the gas mass derived from CO emission lines and dust continuum. The black line denotes the one-to-one relation and the dotted lines show the 0.3\,dex uncertainties. The NICE sample are circles, and the median ratio of -0.18\,dex is indicated by the dashed line. }
\label{fig:CO_Submm}
\end{figure}


We also estimated the gas masses  from CO(3-2) or  CO(4-3) line emission, $M_{\rm gas}^{\rm CO}$, by converting them to the ground-state CO(1-0) luminosity, assuming the CO spectral line energy distribution (CO-SLED) of  star-forming galaxies  at $z$\,=\,2.0\,-\,2.7 from the ASPECS survey \citep{Boogaard2020}. These galaxies exhibit slightly higher CO excitation compared to main sequence galaxies at $z$\,<\,2 \citep{Valentino2020},  but overall agree, within the uncertainties, with the average SLEDs of SMGs at $z$\,=\,1.2\,--\,4.8 \citep{Birkin2021}.
The conversion factor $\alpha_{\rm CO}$ was derived using the mass–SFR–metallicity relation from \citet{Genzel2015} and the metallicity dependence of $\alpha_{\rm CO}$ from \citet{Tacconi2018} for our sample. 

 In Fig.~\ref{fig:CO_Submm} we compare the gas masses estimated using the two methods. The two methods yield consistent results within 0.3\,dex, with $M_{\rm gas}^{\rm RJ}$  approximately 1.5 times higher than $M_{\rm gas}^{\rm CO}$. A similar comparison of CO(1–0) and 1.2\,mm dust continuum measurements in galaxies within the Spiderweb protocluster also shows comparable results, albeit with large scatters \citep{Zhang2024}.
Table~\ref{tab:properties} lists the $\alpha_{\rm CO}$ values and gas masses calculated with different tracers.  For consistency, in the following we adopt gas masses estimated from the dust continuum.

\begingroup
\setlength{\tabcolsep}{3pt} 
\renewcommand{\arraystretch}{1.5} 
\begin{table*}[ht]
\caption{Properties of the  CO or continuum-detected galaxies  in the massive protoclusters of NICE-COSMOS. }
\small
	\centering
	\begin{tabular}{ccccccccccccccccccccccccccccccccccc}
 \hline
 \hline
  Structure & ID & $r/R_{\rm vir}$ & $z_{\rm spec}$ & log$_{10}M_\star$ & SFR  &$\alpha_{\rm CO}$ & log$_{10}M_{\rm gas}^{\rm RJ}$  & log$_{10}M_{\rm gas}^{\rm CO}$ & \textit{SFR}$_{\rm radio}$ &\textit{SFR}$_{\rm 24\mu m}$   \\ 
 & & & & \tiny{[\msun]} &  \tiny{[\msunyr]} & [\scalebox{0.8}{\msun\,(\Kkmspc)$^{-1}$}] & \tiny{[\msun]} & \tiny{[\msun]} & \tiny{[\msunyr]} & \tiny{[\msunyr]}\\ 
 (1)&(2)&(3)&(4)&(5)&(6)&(7)&(8)&(9)&(10)&(11)\\
\hline
  HPC1001   &   ...$^{\ast}$    & 0.07       & 3.613 & 11.2$^{+0.2}_{-0.2}$ & 662$^{+65 }_{-66 }$ &3.59 & 11.48$\pm$0.07 & 11.30$\pm$0.04  & 125$\pm$41  &978$\pm$883  \\  
  HPC1001   & 1272853 & 1.12       & 3.604 & 11.3$^{+0.1}_{-0.2}$ & 496$^{+290}_{-241}$ &3.62 & ...             & 10.87$\pm$0.10  &...           &...           \\       
  HPC1001   & 1274387 & 0.78       & 3.61  & 10.3$^{+0.1}_{-0.1}$ & 68$^{+69 }_{-20 }$  &4.59 & 11.26$\pm$0.18 & 10.84$\pm$0.10  & <159        &...           \\        
\hline
  COS-SBC3  & 1340799 & 0.30       & 2.365 & 11.5$^{+0.1}_{-0.1}$ & 190$^{+90 }_{-61 }$ &3.52 & 11.41$\pm$0.12 & 10.84$\pm$0.07  & <138        &231$\pm$13  \\
\hline
 COS-SBC4    & 840072 & 0.04 & 1.65 &11.4$^{+0.1}_{-0.1}$ & 100$^{+18}_{-18}$ & 3.54 & 11.21$\pm$0.12 &...  & 165$\pm$12        &252$\pm$20  \\
\hline 
  COS-SBC6  & 835289  & 0.07       & 2.323 & 11.1$^{+0.1}_{-0.1}$ & 73$^{+15 }_{-19 }$  &3.62 & 10.95$\pm$0.16 & 10.99$\pm$0.05  & 101$\pm$20  &289$\pm$17  \\ 
  COS-SBC6  & 839791$^{\ast\ast}$&0.10 & 2.324 & 11.0$^{+0.0}_{-0.1}$ & 93$^{+6  }_{-6  }$  &4.01 & 11.51$\pm$0.04 & 11.03$\pm$0.05  &1143$\pm$60  &195$\pm$25  \\  
\hline
  COS-SBCX1 & 1408110 & 0.29       & 2.422 & 11.4$^{+0.1}_{-0.1}$ & 156$^{+30 }_{-31 }$ &3.54 & 11.07$\pm$0.15 & 11.03$\pm$0.05  & 134$\pm$16  &183$\pm$33  \\  
\hline
  COS-SBCX3 & 1088927 & 0.04       & 3.032 & 10.5$^{+0.1}_{-0.2}$ & 100$^{+45 }_{-90 }$ &4.78 & 11.28$\pm$0.15 & 11.09$\pm$0.05  & 495$\pm$56  &472$\pm$76   \\   
  COS-SBCX3 & 1088787 & 0.11       & 3.03  & 10.7$^{+0.1}_{-0.1}$ & 53$^{+40 }_{-12 }$  &4.31 & 11.17$\pm$0.19 & 11.25$\pm$0.04  & 490$\pm$57  &284$\pm$114  \\   
\hline
  COS-SBCX4 & 1049510 & 0.07       & 2.642 & 11.1$^{+0.1}_{-0.1}$ & 257$^{+83 }_{-186}$ &3.62 & 10.94$\pm$0.22 & 10.78$\pm$0.07  & <123        &137$\pm$31  \\   
  COS-SBCX4 & 1050531 & 0.05       & 2.647 & 11.2$^{+0.1}_{-0.1}$ & 416$^{+135}_{-265}$ &3.69 & 11.66$\pm$0.04 & 11.43$\pm$0.02  & 533$\pm$37  &496$\pm$51  \\   
  COS-SBCX4 & 1049929 & 0.14       & 2.648 & 10.9$^{+0.2}_{-0.2}$ & 195$^{+120}_{-101}$ &3.78 & 10.80$\pm$0.31 & 10.67$\pm$0.07  & <141        &...          \\  
\hline
  COS-SBCX7 & 392639  & 0.75       & 2.413 & 10.3$^{+0.1}_{-0.1}$ & 114$^{+25 }_{-112}$ &4.66 & ...              & 11.00$\pm$0.12  & <84         &36$\pm$17   \\  
  COS-SBCX7 & 394944  & 0.14       & 2.416 & 11.1$^{+0.1}_{-0.1}$ & 92$^{+70 }_{-25 }$  &3.60 & 11.22$\pm$0.12 & 10.89$\pm$0.06  & <87         &184$\pm$16  \\  
  COS-SBCX7 & 394609  & 0.01       & 2.416 & 10.8$^{+0.1}_{-0.1}$ & 14$^{+11 }_{-4  }$  &3.93 & 11.24$\pm$0.11 & 11.03$\pm$0.05  & 186$\pm$24  &239$\pm$17  \\  
  COS-SBCX7 & 392257  & 0.15       & 2.415 & 10.7$^{+0.1}_{-0.1}$ & 91$^{+61 }_{-49 }$  &4.10 & 10.81$\pm$0.34 & 10.86$\pm$0.07  & 212$\pm$25  &122$\pm$17  \\  
\hline
	\end{tabular}
 \begin{tablenotes}
      \small
      \item \textbf{Notes.} 
      \textit{Columns}: (1)--(6) protocluster ID, galaxy ID in COSMOS2020 \citep{Weaver2022}, distance to protocluster center, redshift, stellar mass, and SFR as given in \citet{Sillassen2024}. (7) CO-to-H$_2$ conversion factor. (8) \& (9) Molecular gas masses from dust continuum and from CO. (10) SFR from 3GHz observations, with 3$\sigma$ upper limits for nondetections. (11) SFR from 24$\mu$m fluxes adopting conversion factors from \citet{Wuyts2008}. 
       For details, see Section~\ref{sec:data}. $^{\ast}$: This source could be a blend of several galaxies \citep{Sillassen2022}. $^{\ast\ast}$: Radio-loud galaxy.
    \end{tablenotes}
	\label{tab:properties}
\end{table*}
\endgroup

\section{Star formation and cold gas properties}
\subsection{Enhanced Star Formation in Massive Halos in the Early Universe}

\begin{figure}[ht]
\centering
\includegraphics[align=c, width=\linewidth]{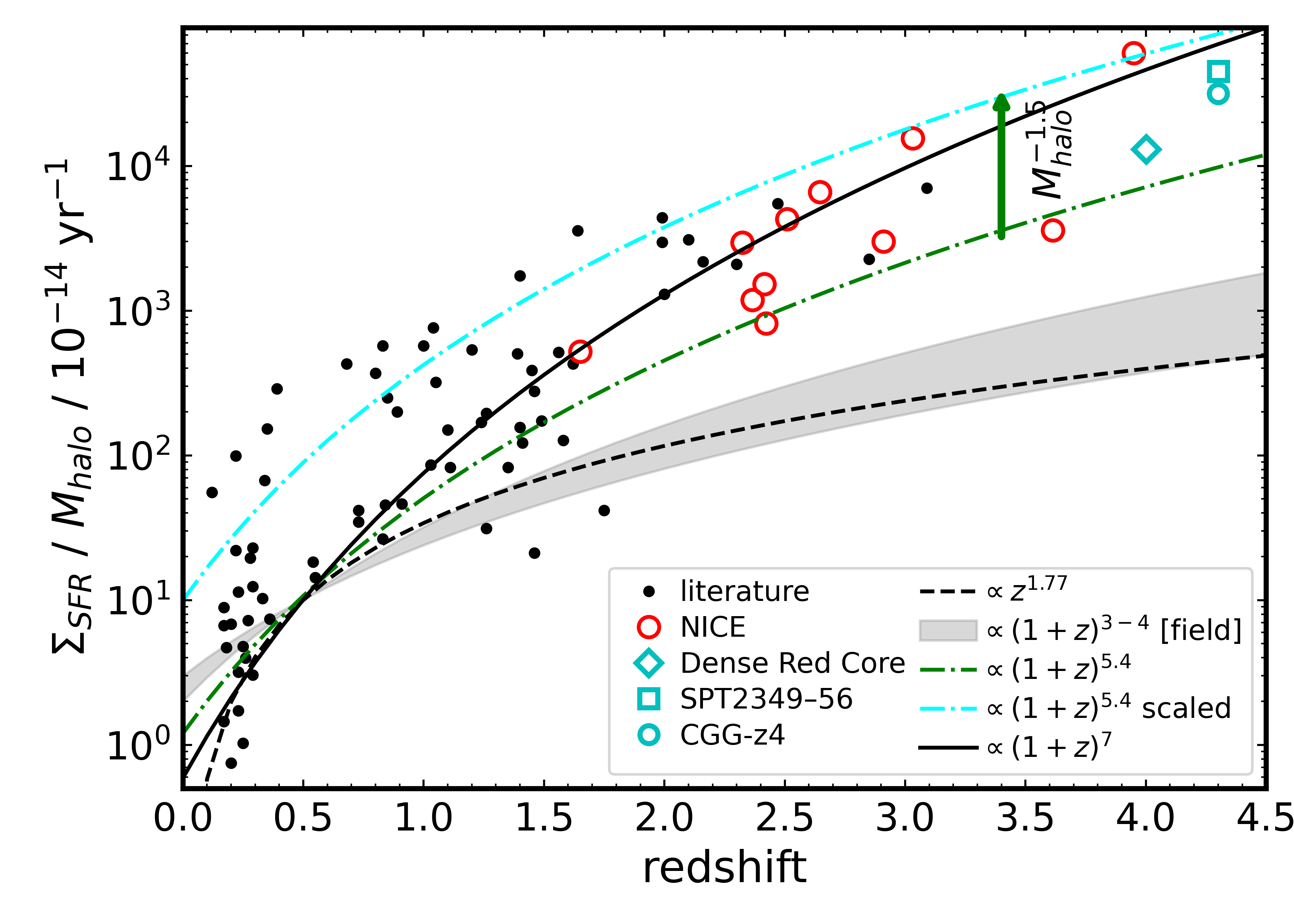}
\caption{Total SFR per unit halo mass as a function of redshift. Shown as red circles are the eight massive protoclusters in the COSMOS field and  LH-SBC3 at $z$\,=\,3.95, the most distant protocluster identified in the NICE survey \citep{Zhou2024}, as well as   two other prototypes within the NICE survey, RO1001 at $z$\,=\,2.91 \citep{Kalita2021} and CL~J1001 at $z$\,=\,2.51 \citep{Wang2018}. Two other starbursting protoclusters, SPT2349-56 at $z$\,=\,4.3  \citep{Hill2022} and Distant Red Core at $z$\,=\,4.0 \citep{Long2020}, are shown as a cyan square and diamond, respectively. CGG-z4,  a compact galaxy group at $z$\,=\,4.3 in the COSMOS field \citep{Brinch2025}, is  shown as a cyan circle. The  black dots and the curves show the  (proto)clusters and redshift evolution  compiled in \citet{Alberts2022}.  
 The (proto)cluster data are drawn from \citet{Haines2013, Santos2014, Santos2015, Popesso2015, Popesso2015b, Ma2015, Alberts2014, Alberts2016, Casey2016, Strazzullo2018, Smith2019, Lacaille2019}. The black dashed line shows a redshift evolution trend $\propto z^{1.77}$ \citep{Popesso2011}, while the green dash-dotted line indicates a $\propto (1+z)^{5.4}$ relation \citep{Alberts2014, Webb2013, Bai2009}. The cyan dash-dotted line also follows the $\propto (1+z)^{5.4}$ trend, but is scaled from logM$_{\rm 200}$/\msun\,=\,14.5 to 14,  assuming $\Sigma$SFR/$M_{\rm halo} \propto M_{\rm halo}^{-1.5}$ \citep{Webb2013}. The solid line shows a steeper evolution of $\propto (1+z)^7$  \citep{Cowie2004, Geach2006, Biviano2011}, normalized at $z$\,=\,0.5. For field galaxies, the gray shaded region represents the redshift evolution trend of $\propto (1+z)^{3-4}$ based on \citet{LeFloch2005, Rujopakarn2010, Sargent2012, Ilbert2015}.
}
\label{fig:alberts}
\end{figure}

The value of  $\Sigma_{\rm SFR} /M_{\rm halo}$ in clusters increases with redshift, following an empirical relation of (1+$z$)$^n$ (see \citealt{Alberts2022} for a review). The NICE survey extends such studies to $z$>2.
We adopted the integrated SFR and best estimates of halo masses of the eight protoclusters from \citep{Sillassen2024} to obtain the total SFR per unit halo mass ($\Sigma_{\rm SFR} /M_{\rm halo}$). The SFR is from the FIR spectral energy distribution (SED) fitting using the dust templates in \citet{Magdis2012}, and the halo masses are the average results from six methods including stellar mass-to-halo mass relations, overdensity with galaxy bias, and NFW profile fitting to radial stellar mass density. 
 In Fig.~\ref{fig:alberts} we include the eight massive protoclusters in NICE-COSMOS;  LH-SBC3 at $z$\,=\,3.95, which is the most distant protocluster identified in the NICE survey \citep{Zhou2024};   protoclusters RO1001 at $z$\,=\,2.91 \citep{Kalita2021} and CL~J1001 at $z$\,=\,2.51 \citep{Wang2018},  which are prototypes within the NICE survey; and SPT2349-56 at $z$\,=\,4.3  \citep{Hill2022} and Distant Red Core at $z$\,=\,4.0 \citep{Long2020}, which are starbursting protoclusters  similar to the NICE clusters. 
We found that $\Sigma_{\rm SFR} /M_{\rm halo}$ increases by an order of magnitude from $z$\,=\,2 to $z$\,=\,4,  following the evolutionary trend observed in galaxy clusters up to $z$\,=\,1  \citep[$\Sigma_{\rm SFR} /M_{\rm halo}$\,$\propto$\,(1+$z$)$^{5.4}$, ][]{Webb2013}, which has recently been shown to extend to clusters at $z$\,=\,1.6--2 \citep{Smail2024}.  These values are one to two orders of magnitude higher than those observed in the field \citep[$\propto$\,(1+$z$)$^{3-4}$][]{Sargent2012, Ilbert2013, Ilbert2015}. This may be  due to the evolution of the SFMS, modulated by the evolution of the $M_{\rm \star, tot}/M_{\rm halo}$ relation, and the reduction of the quenched fraction (Elias et al. in prep.).

\subsection{Star-forming main sequence in dense environments}

\begin{figure}[ht]
\centering
\includegraphics[align=c, width=0.9\linewidth]{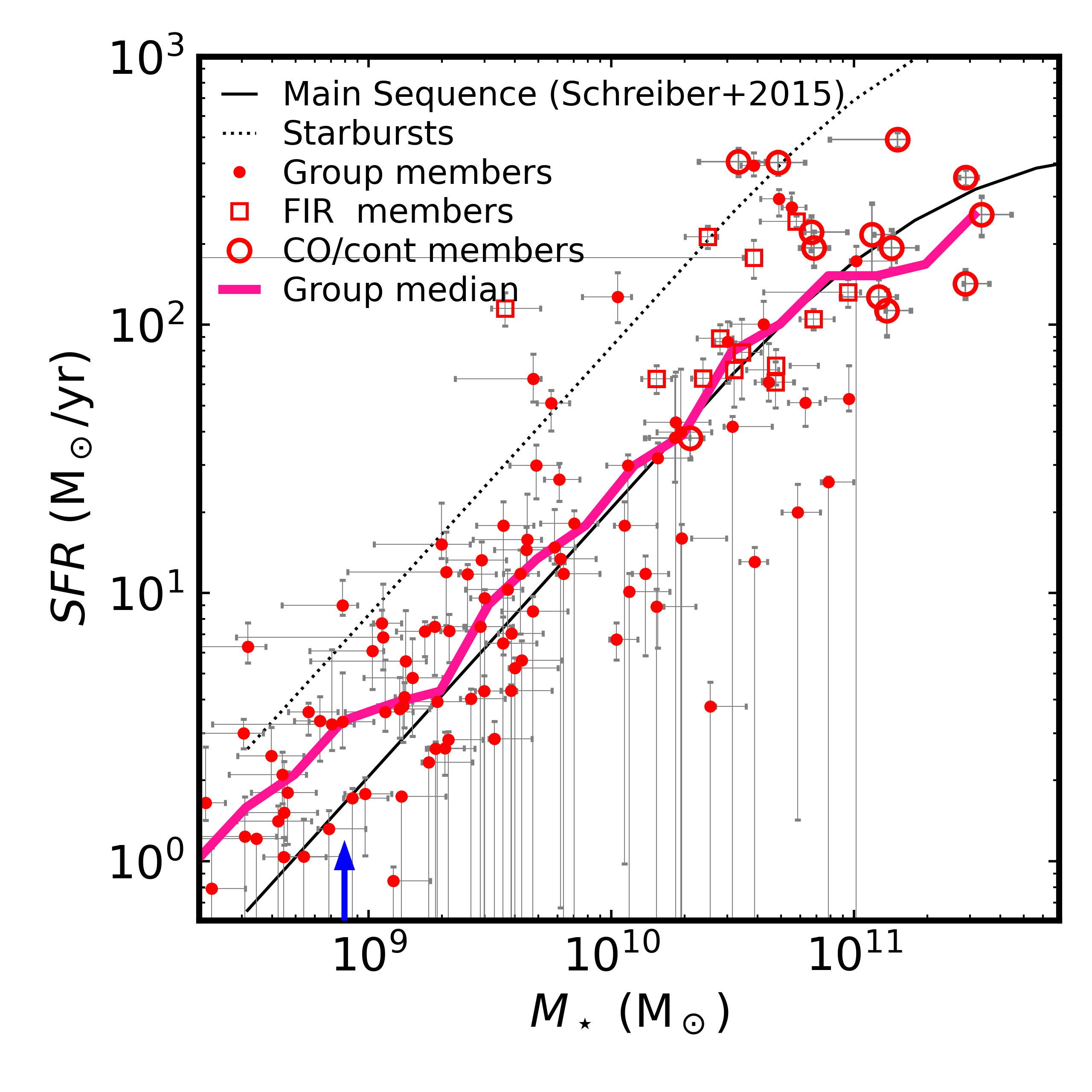}
\caption{
Star-forming main sequence of galaxy members in the eight protoclusters within NICE-COSMOS. CO-detected and dust continuum-detected members are highlighted with open red circles, while other FIR-detected galaxies are marked with open red squares. The remaining cluster members are represented as red dots. To enable comparison, the galaxies are scaled to a common redshift of $z$\,=\,2.5, using the SFMS normalization and redshift evolution from \citet{Schreiber2015}. The SFRs of FIR-detected members are derived from 3\,GHz radio emission or 24$\mu$m fluxes, while the SFRs for the rest are obtained from COSMOS2020 (see Section~\ref{sec:mstar_sfr} for details). The pink curve represents the median SFR of member galaxies within the NICE-COSMOS protoclusters. For reference, the black solid curve shows the main sequence at $z = 2.5$ from \citet{Schreiber2015}, and the dotted curve indicates the starburst regime, defined as \rsb\,=\,4 \citep{Rodighiero2011}. The blue arrow indicates the mass completeness limit of star-forming galaxies at $z$\,=2.5 from COSMOS2020 \citep{Weaver2022}.
} 
\label{fig:MS}
\end{figure}

The SFR and stellar mass of all \textit{UVJ}-active cluster members \citep{Williams2009} are shown in Fig.~\ref{fig:MS}.  Galaxies at different redshifts are scaled to $z$\,=\,2.5, following the SFMS normalization and redshift evolution from \citet{Schreiber2015}. Despite significant scatter, the median values of these galaxies closely follow the main sequence in the field,  in line with the results from H$\alpha$ emitters in  protoclusters  at $z\sim$\,2.5 \citep[e.g., Spiderweb,][]{Perez-Martinez2023}. 

The CO-detected members, which are among the most massive with the highest SFRs, may indicate a greater abundance of gas fueling star formation in these galaxies. However, this trend is not observed in the CO-detected H$\alpha$ emitters in the Spiderweb protocluster \citep{Perez-Martinez2025}, where SFRs are derived from the dust-corrected H$\alpha$ fluxes, possibly due to the differences in protocluster selection criteria.  In contrast, the less massive galaxies display slightly elevated star formation compared to field galaxies, a trend similar to that observed in H$\alpha$ emitters within the young protocluster USS1558-003 at $z$\,=\,2.53 \citep{Hayashi2012, Daikuhara2024}.  

It is important to note that the cluster members were primarily selected  based on photometric redshifts, which may include interlopers. This may affect the results shown in Fig.~\ref{fig:MS}, especially at the low-mass end, where the contamination is  most significant. On the other hand, the high-mass end is less impacted as it is dominated by CO-detected galaxies that are spectroscopically confirmed members.

\begin{figure}[ht]
\centering
\includegraphics[align=c, width=0.9\linewidth]{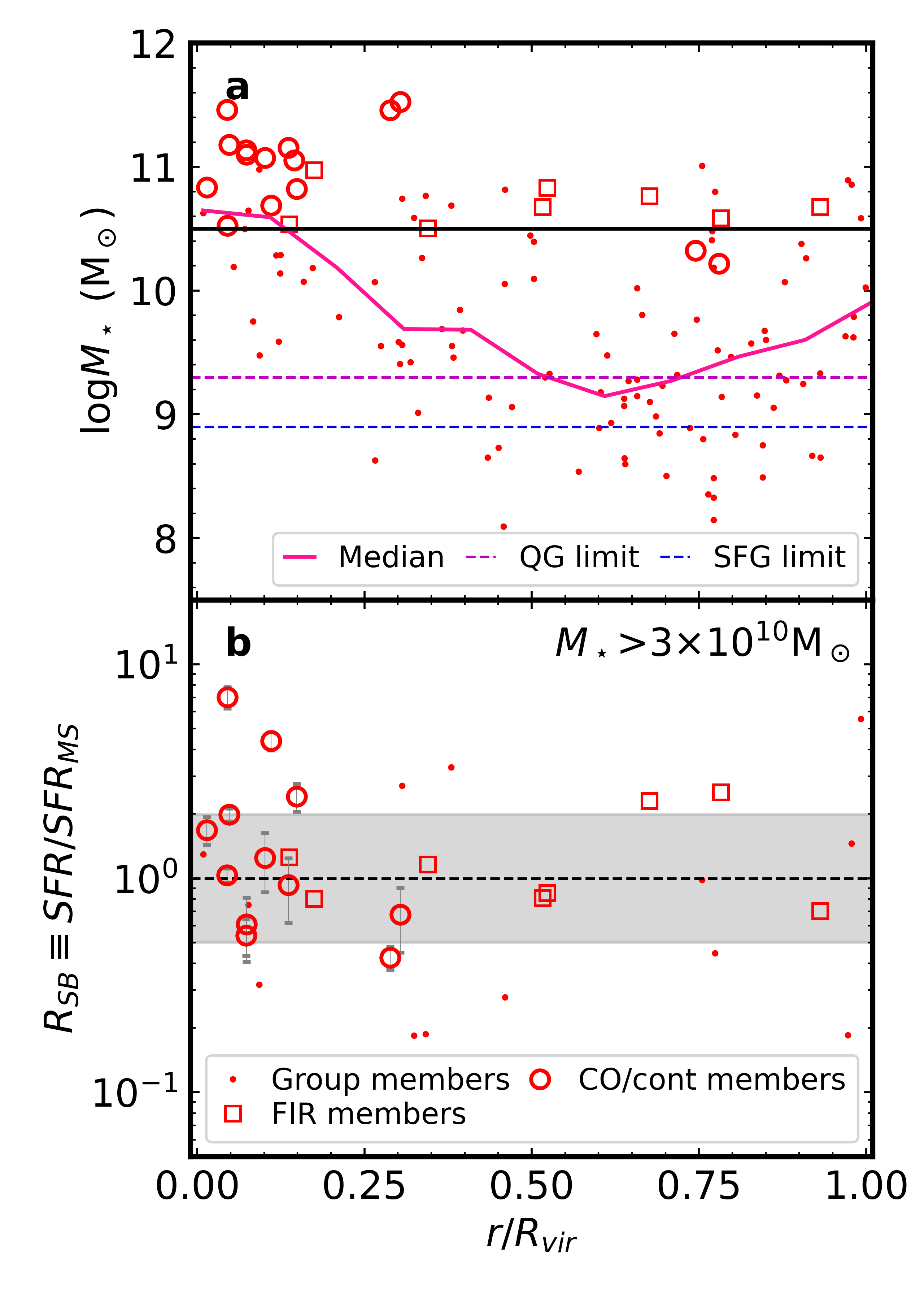}
\caption{\textbf{Panel a:} Stellar mass as a function of  distance to the cluster centers. The symbols are as defined in Fig.~\ref{fig:MS}. The completeness limits of star-forming galaxies and quiescent galaxies are also shown as references. \textbf{Panel b:} Starburstiness (\rsb) of  massive galaxies (\mstar\,>\,3$\times$10$^{10}$\,\msun) as a function of the distance to the cluster center. The dashed line shows the field main sequence and the shaded region shows the uncertainty.
}
\label{fig:SB}
\end{figure}


\subsection{Build-up of massive galaxies in the dense cores} 

We analyzed the stellar mass distribution of the cluster galaxies and found that more massive galaxies are located closer to the protocluster center (Fig.~\ref{fig:SB}a). The median stellar mass  within 0.1\,$R_{vir}$ is $\sim$10$^{10.6}$\msun, compared to $\sim$10$^{9.5}$\msun\, at the outskirts, representing more than 1 dex difference. This difference in the mass of  main sequence galaxies naturally  enhances the overall star formation activities in these protoclusters. 
Out of the 31  massive galaxies (\mstar\,>\,3$\times$10$^{10}$\,\msun), 21 are located within 0.3$R_{vir}$, with all but one showing detected CO emission. This aligns with the high fraction of SMGs predicted in protocluster cores by  \citet{Araya-Araya2024}  simulations and highlights the presence of substantial gas reservoirs in these regions.  Similarly, a top-heavy stellar mass function has been identified in the starbursting protocluster CL~J1001 \citep{Sun2024}, suggesting a prevalence of massive galaxy accumulation in such protoclusters.
Although these massive galaxies generally follow the main sequence, those in the cores are on average situated above the main sequence compared to their counterparts in the outskirts (Fig.~\ref{fig:SB}b).

The field of view (FoV) of the NOEMA and ALMA observations, with a  full width at half maximum (FWHM) of $\sim$30\arcsec and $\sim$1\arcmin, respectively,  is sufficiently large to encompass the area within the virial radius of the protoclusters.  This indicates that we are witnessing the rapid build-up of massive galaxies in the cores of these protoclusters at cosmic noon.

\subsection{Gas content}

\begin{figure*}[ht]
\centering
\includegraphics[align=c, width= 0.9\linewidth]{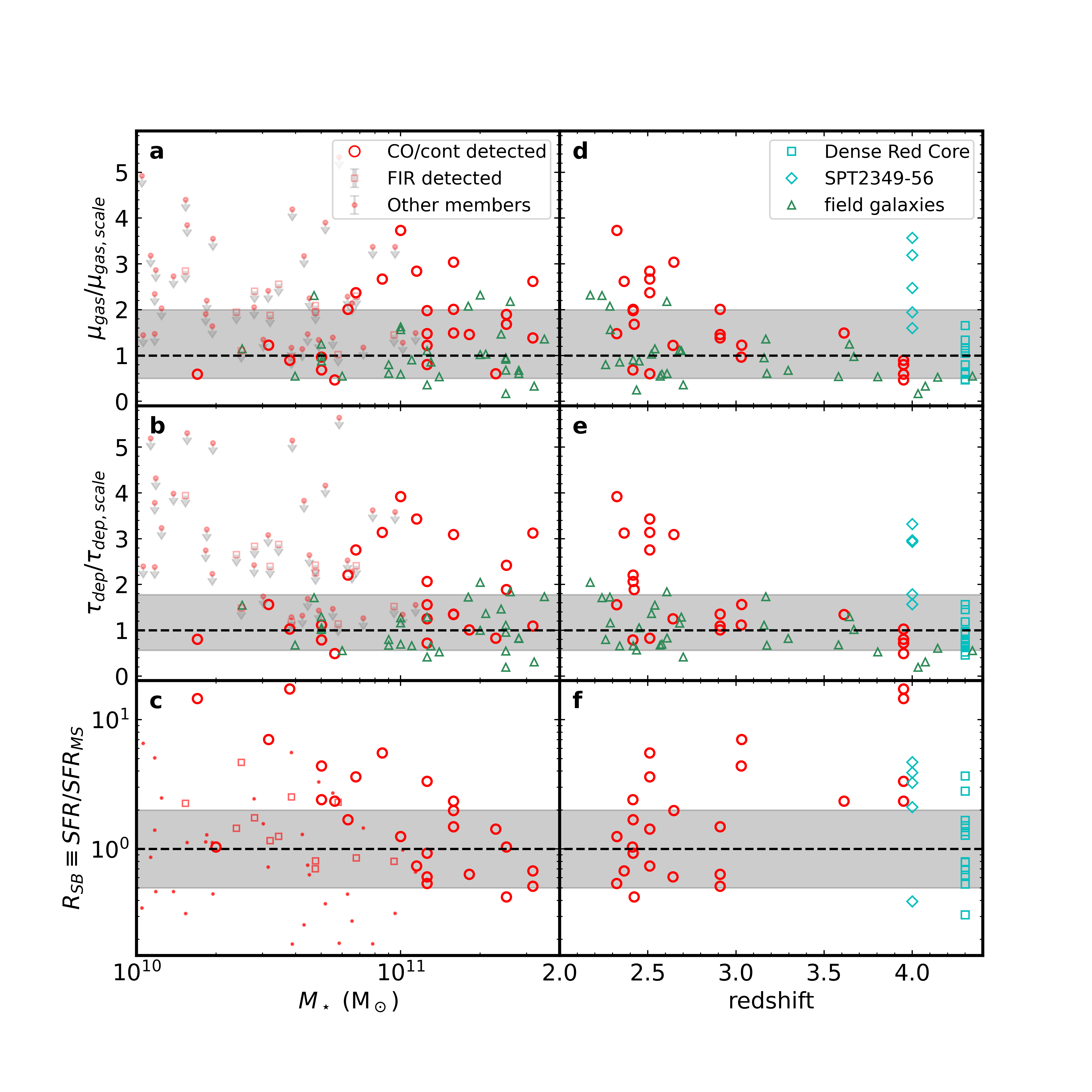}
\caption{Gas-to-stellar mass ratio (\fgas = \mgas/\mstar), gas depletion time ($\tau_{gas}=M_{gas}/SFR$), and starburstiness (\rsb) as a function of stellar mass (left column) and redshift (right column). \fgas\,   and \taugas\, are scaled relative to field galaxy levels (\fgas$_{,\rm scale}$, \taugas$_{,\rm scale}$), accounting for the stellar masses, redshifts, and starburstiness of each galaxy. The red circles represent galaxies detected in CO or dust continuum in the NICE protoclusters, including the eight in the COSMOS field \citep{Sillassen2024}, LH-SBC3 \citep{Zhou2024}, RO1001 \citep{Kalita2021}, and CL J1001 \citep{Wang2018}, while red squares and dots indicate the 3$\sigma$ upper limits of those detected in the FIR and the rest of the cluster galaxies. Two additional starbursting protoclusters at $z\gtrsim$\,4, SPT2349-56 \citep{Hill2022} and
Distant Red Core \citep{Long2020}, are shown in cyan. Field galaxies from \citet{Kaasinen2019, Boogaard2020, FriasCastillo2023} are represented as green triangles,  with the molecular gas content derived from CO(1-0) emission lines. The black dashed lines denote the positions of field galaxies; the gray shading indicates the uncertainties. }
\label{fig:gas}
\end{figure*}

The gas masses of CO-detected NICE-COSMOS galaxies span the range \mgas\,=\,10$^{10.6}$\,$\sim$\,10$^{11.7}$\,\msun. The gas-to-stellar mass ratio \fgas\,=\,\mgas/\mstar\, and gas depletion time \taugas\,=\,\mgas/SFR of NICE protocluster galaxies have median values of \fgas$_{\rm,NICE}$\,=\,\fgasMed  and \taugas$_{\rm,NICE}$\,=\,\taugasMed\,Gyr. These values are consistent with those observed in other protocluster galaxies at similar redshifts  \citep[e.g.,][]{Tadaki2019, Perez-Martinez2025}. However, protoclusters at higher redshifts ($z$\,$\gtrsim$\,4), such as SPT2349-56 \citep{Hill2022} and LH-SBC3 \citep{Zhou2024}, tend to show lower \fgas\, and \taugas, while massive galaxies in Distant Red Core at $z$=4 \citep{Long2020} have comparable gas content to those in protoclusters at $z$\,$\sim$\,2.5.  For a fair comparison, we recalculated the gas masses of galaxies in the literature using the same method described in Section~\ref{sec:gasmass}  when CO(1-0) measurements were unavailable. 

To assess the environmental impact on the gas content, we normalized \fgas\, and \taugas\, to the corresponding field levels,  accounting for their dependence  on stellar mass, starburstiness, and redshift. This analysis utilized parameterized relations compiled from the \acosmos\, project, based on a systematic mining of ALMA archival data for CO-observed galaxies in the COSMOS field \citep{Liu2019}.  The results are presented in Fig.~\ref{fig:gas}.

The 3$\sigma$ detection limit of the NICE observations corresponds to a gas mass of \mgas\,$\sim$\,10$^{11}$\,M$_\odot$ for galaxies at $z$\,$\sim$\,2.5, meaning that our sample predominantly includes the most massive members and the less massive galaxies with high starburstiness (Fig.~\ref{fig:gas}c). The most massive ones (\mstar\,>\,10$^{10.8}$\,M$_\odot$), which are  concentrated in the core regions, are nearly all detected via CO or dust continuum emission. These massive galaxies generally exhibit higher gas content ($\sim$2$\times$) compared to  their counterparts in the field (Fig.~\ref{fig:gas}\,a, b). 
Several of them display significantly elevated gas-to-stellar mass ratios and gas depletion times relative to  field galaxies. However, their SFRs are not correspondingly enhanced, suggesting that they have abundant gas reservoirs to sustain their growth or their SFRs are prevented from increasing to starburst levels due to  merger-related processes or negative feedback. 

We also note that the field galaxies \citep{Kaasinen2019, Boogaard2020, FriasCastillo2023}, whose molecular gas content is derived from CO(1–0) emission adopting $\alpha_{\rm CO}$ obtained using the method outlined in Section~\ref{sec:gasmass}, fall within the field-level range defined by the \acosmos\, project.
This is consistent with the findings from  \citet{Calvi2023}, who found a correlation between molecular gas masses of SMGs and the significance of the associated overdensity.

Most galaxies with lower stellar masses are not detected in CO or dust continuum, except for the starbursts. However, these starburst galaxies exhibit gas-to-stellar mass ratios and gas depletion times comparable to those of field galaxies. It remains unclear whether the CO- or dust-nondetected cluster members have enhanced or deficient gas content. 

In contrast, three protoclusters traced by Ly$\alpha$ emitters and/or red galaxies surrounding a radio galaxy \citep{Tadaki2019} show the opposite trend, with a declining gas fraction and gas depletion time as stellar mass increases. This indicates a mass-dependent environmental effect, where gas accretion is accelerated in less massive galaxies, while suppressed in more massive ones. These contrasting trends may reflect different evolutionary stages of the two protocluster populations.

Interestingly, we find that most of the gas-rich galaxies belong to protoclusters at $z$\,$\sim$\,2.5, whereas massive galaxies in more distant protoclusters exhibit a declining gas content, though still generally within the uncertainties of field galaxies (Fig.~\ref{fig:gas}d,e),  approximately 2 times  at $z$\,=\,2.5 and 1.2 times at $z$\,=\,4.  Similarly, \citet{Calvi2023} identified a possible transitioning phase at $z$\,$\sim$\,4, with $z$\,<\,4 protoclusters more populated by dusty galaxies. We speculate that our observed trend  may partly reflect a selection bias in constructing the field sample, which tends to favor the brightest sources at high redshifts, typically located in dense environments.

We further checked the overall  mean intensity of the radiation field, <$U$>, of the NICE-COSMOS protoclusters, as measured from the integrated FIR SED fitting in \citet{Sillassen2024}. <$U$> can be used to infer the metallicity weighted star formation efficiency (SFE\,$\equiv$1/\taugas) and is proportional to dust temperature \citep{Magdis2012, Magdis2017}. The evolution of <$U$> with redshift for the eight protoclusters generally follows the trend observed in main sequence galaxies \citep{Bethermin2015}; a few of them show lower <$U$>. This is in line with the moderately longer gas depletion time observed in the individual member galaxies and may indicate a colder dust temperature in these protoclusters. 

\begin{figure}[ht]
\centering
\includegraphics[align=c, width=0.9\linewidth]{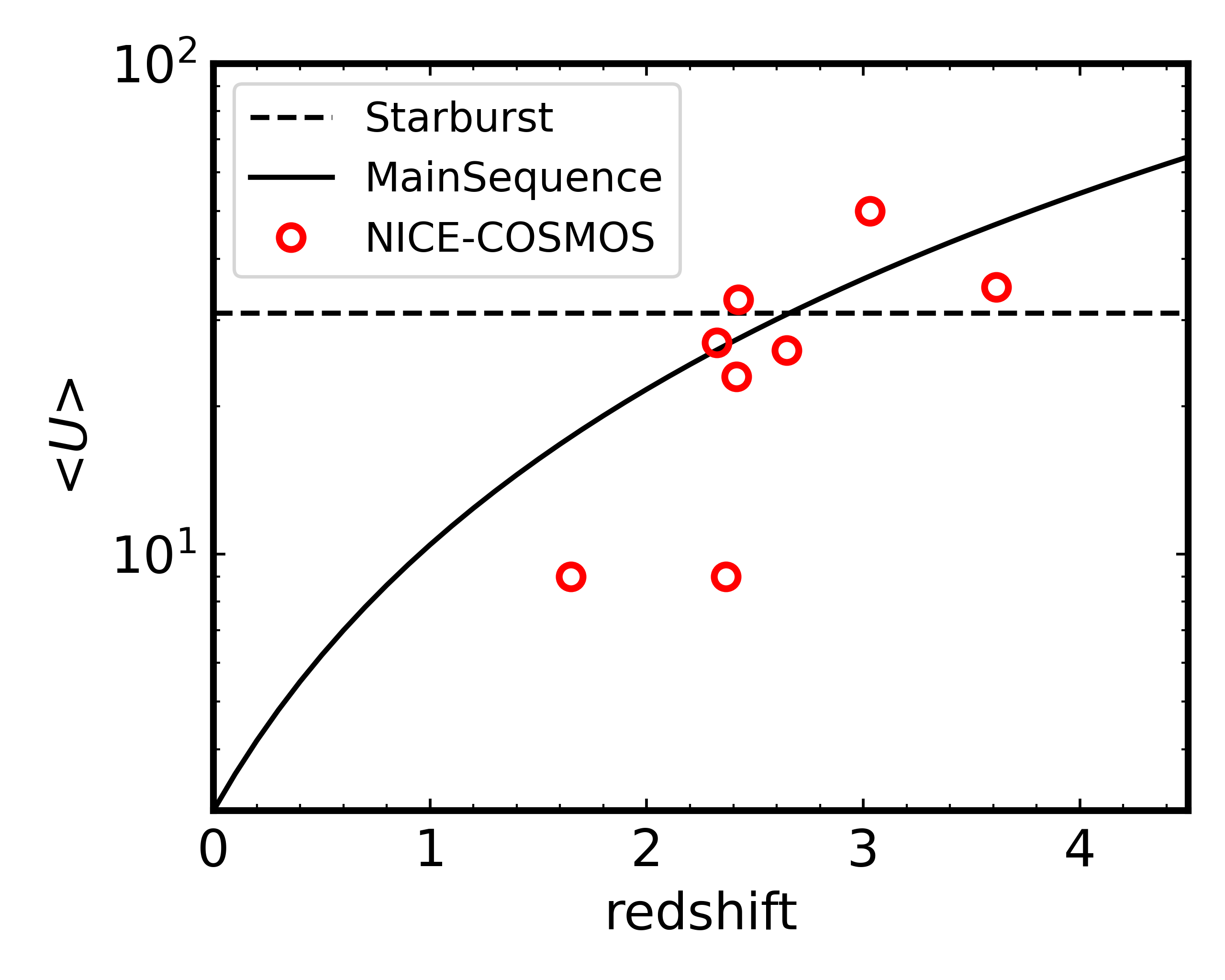}
\caption{ Evolution of the mean intensity of the radiation field, <$U$>, with redshift. The red circles show the overall <$U$> of the eight protoclusters in NICE-COSMOS. The solid curve and dashed line represent the trend of main sequence galaxies and starbursts, as in \citet{Bethermin2015}.
}
\label{fig:Uz}
\end{figure}


\section{Summary and discussion}

In this paper we analyzed the star formation and cold gas properties of eight protoclusters in the COSMOS field, selected through the NICE survey, which targets intensively star-forming protocluster cores at cosmic noon.

We find that the total SFR per unit halo mass ($\Sigma_{\rm SFR} /M_{\rm halo}$) in protoclusters at $z$\,>\,2 increases rapidly with redshift, scaling as $(1+z)^{5.4}$ (Fig.~\ref{fig:alberts}). 
This could be due to either a higher number of massive star-forming galaxies in massive halos or enhanced star formation in cluster galaxies. 
Additionally, we find that the fraction of quiescent galaxies decreases relative to that observed in lower-redshift (proto)clusters,  and that the streaming mass of the halos (i.e., the threshold mass delimiting the hot and cold in hot accretion regimes; see \citealt{Daddi2022} and \citealt{Dekel2009}) is also a good predictor for the quiescent fraction, making the gas accretion mode a potential explanation for the quiescent fraction evolution across halos (Elias et al. in prep.).
We also caution that the higher $\Sigma_{\rm SFR} /M_{\rm halo}$ values at higher redshifts could also be partly driven by lower halo masses.  

Nevertheless, this rise in star formation activity in protoclusters is not reflected in individual member galaxies. The SFMS remains comparable to that observed in the field (Fig.~\ref{fig:MS}), suggesting that the elevated  $\Sigma_{\rm SFR} /M_{\rm halo}$  in protoclusters is not driven by a higher fraction of starburst galaxies in dense environments.  We note that the cluster members may be affected by contamination due to photometric redshift uncertainties \citep{Sillassen2024}. However, the massive end is dominated by CO-detected and thus spectroscopically confirmed members, making the main sequence at the high-mass end less susceptible to interlopers.

Using observations from NOEMA and ALMA in the NICE survey, we investigated the gas content of cluster galaxies. The CO or dust continuum-detected galaxies are predominantly massive cluster members (\mstar\,>\,3$\times$10$^{10}$\,\msun), which preferentially reside in the  core of the protoclusters ($r\lesssim$\,0.3$R_{vir}$, Fig.~\ref{fig:SB}). This indicates an overabundance of massive star-forming galaxies in these protocluster cores,  which contributes to their elevated $\Sigma_{\rm SFR} /M_{\rm halo}$. 
This finding aligns with the simulations by \citet{Araya-Araya2024}, which identify an excess of massive star-forming galaxies undergoing a submillimeter-bright phase in protocluster cores, suggesting a downsizing formation scenario for massive galaxies.

In terms of stellar-to-gas mass ratio and gas depletion time,   the median values are consistent with field levels. However, the most massive galaxies (\mstar\,>\,8$\times$10$^{10}$\,\msun) are notably more gas-rich than their field counterparts (Fig.~\ref{fig:gas}). This suggests efficient cold gas accretion in the central regions of these protoclusters, supporting the growth of these galaxies to higher masses while remaining in the main sequence mode, without triggering starburst phases. 
For the less massive galaxies,  CO or dust continuum detections are limited to the most starbursting ones, which follow the scaling relation \citep[\fgas($z$, \mstar, \rsb), \taugas($z$, \mstar, \rsb), ][]{Liu2019} observed in the field galaxies.   The moderately higher \taugas\, is consistent with the overall mean intensity of the radiation field in the protoclusters, indicating a generally cold dust temperature in these dense environments.
 However, we caution that the NICE survey is biased toward protoclusters in active star-forming phases. The influence of such environments on member galaxies may differ from that observed in protoclusters selected through other methods.

\begin{acknowledgements}

 We thank the anonymous referee for  valuable comments.
This work was supported by National Natural Science Foundation of China (Project Nos. 13001103, 12173017 and Key Project No.12141301), National Key R\&D Program of China (grant no. 2023YFA1605600), Scientific Research Innovation Capability Support Project for Young Faculty (Project No. ZYGXQNJSKYCXNLZCXM-P3),  and the China Manned Space Program with grant no. CMS-CSST-2025-A04.
L.Z. and Y.S. acknowledges the support from the National Key R\&D Program of China No. 2022YFF0503401 and 2018YFA0404502, the National Natural Science Foundation of China (NSFC grants 12141301, 12121003, 11825302). Y.S. thanks the support by the New Cornerstone Science Foundation through the XPLORER PRIZE.
Y.J.W. acknowledges support by National Natural Science Foundation of China (Project No. 12403019) and Jiangsu Natural Science Foundation (Project No. BK20241188).
GEM acknowledges the Villum Fonden research grant 13160 “Gas to stars, stars to dust: tracing star formation across cosmic time,” grant 37440, “The Hidden Cosmos,” and the Cosmic Dawn Center of Excellence funded by the Danish National Research Foundation under the grant No. 140. 
CGG acknowledges support from CNES.
ZJ  acknowledges funding from JWST/NIRCam contract to the University of Arizona NAS5-02015.
ID acknowledges funding by the INAF Minigrant "Harnessing the power of VLBA toward a census of AGN and star formation at high redshift" and by the European Union – NextGenerationEU, RRF M4C2 1.1, Project 2022JZJBHM: "AGN-sCAN: zooming-in on the AGN-galaxy connection since the cosmic noon" - CUP C53D23001120006.
CdE acknowledges funding from the MCIN/AEI (Spain) and the “NextGenerationEU”/PRTR (European Union) through the Juan de la Cierva-Formación program (FJC2021-047307-I).
SJ is supported by the European Union's Horizon Europe research and innovation program under the Marie Sk\l{}odowska-Curie grant agreement No. 101060888.
This research used APLpy, an open-source plotting package for Python \citep{aplpy2012, aplpy2019}.
This work is based on observations carried out under project number M21AA with the IRAM NOEMA Interferometer. IRAM is supported by INSU/CNRS (France), MPG (Germany) and IGN (Spain).
This research uses data obtained  through the Telescope Access
Program (TAP), which is funded by the National Astronomical Observatories, Chinese Academy of Sciences, and the Special Fund for Astronomy from the Ministry of Finance.
This work is based on observations obtained with WIRCam, a joint project of CFHT, Taiwan, Korea, Canada, France, at the Canada-France-Hawaii Telescope (CFHT) which is operated by the National Research Council (NRC) of Canada, the Institute National des Sciences de l'Univers of the Centre National de la Recherche Scientifique of France, and the University of Hawaii.
\end{acknowledgements}

\bibliographystyle{aa-yang} 
\bibliography{bibitem.bib} 
%

\appendix
\onecolumn 
\section{NICE-COSMOS images}
\begin{figure}[H]
\centering
\includegraphics[width=3.6in]{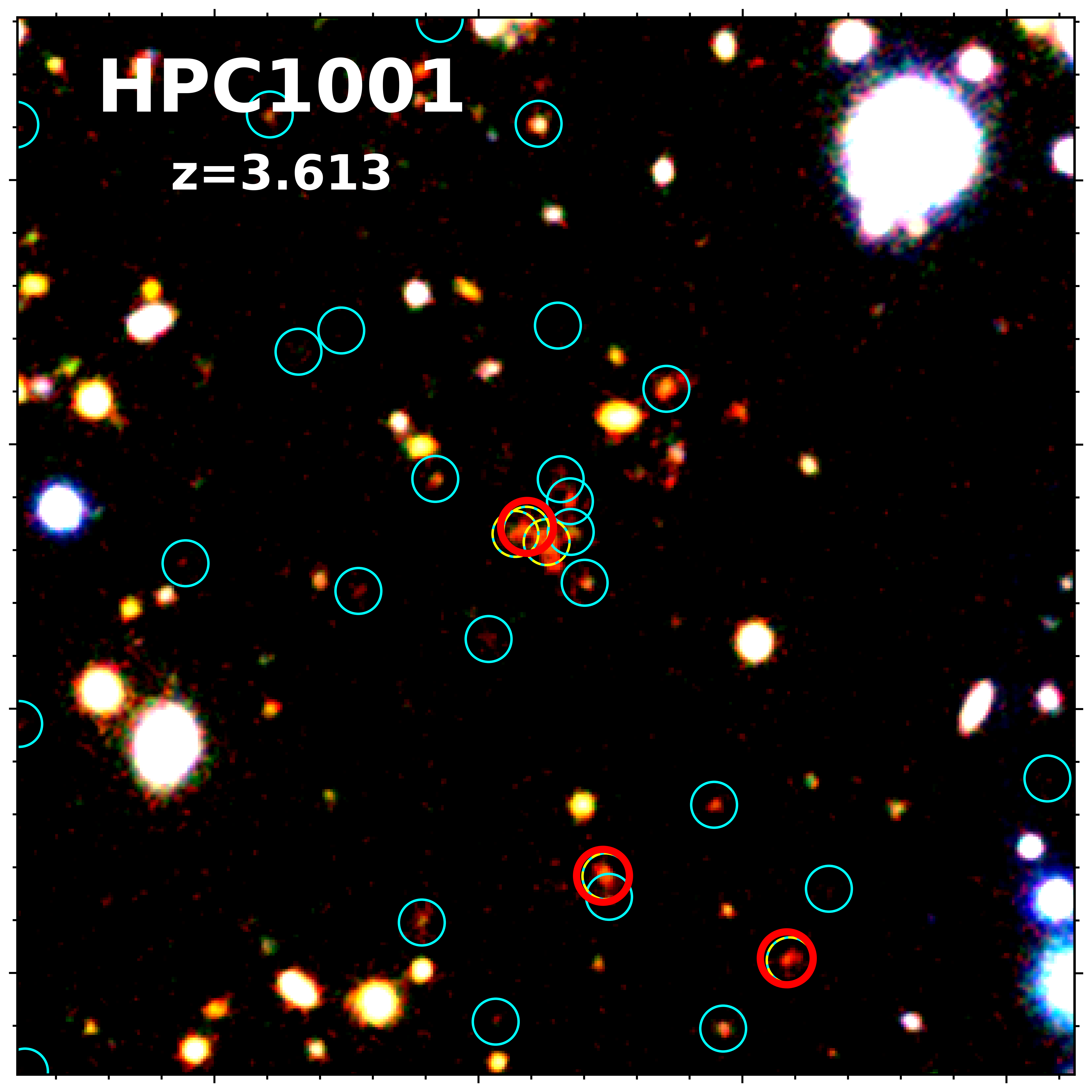}
\includegraphics[width=3.6in]{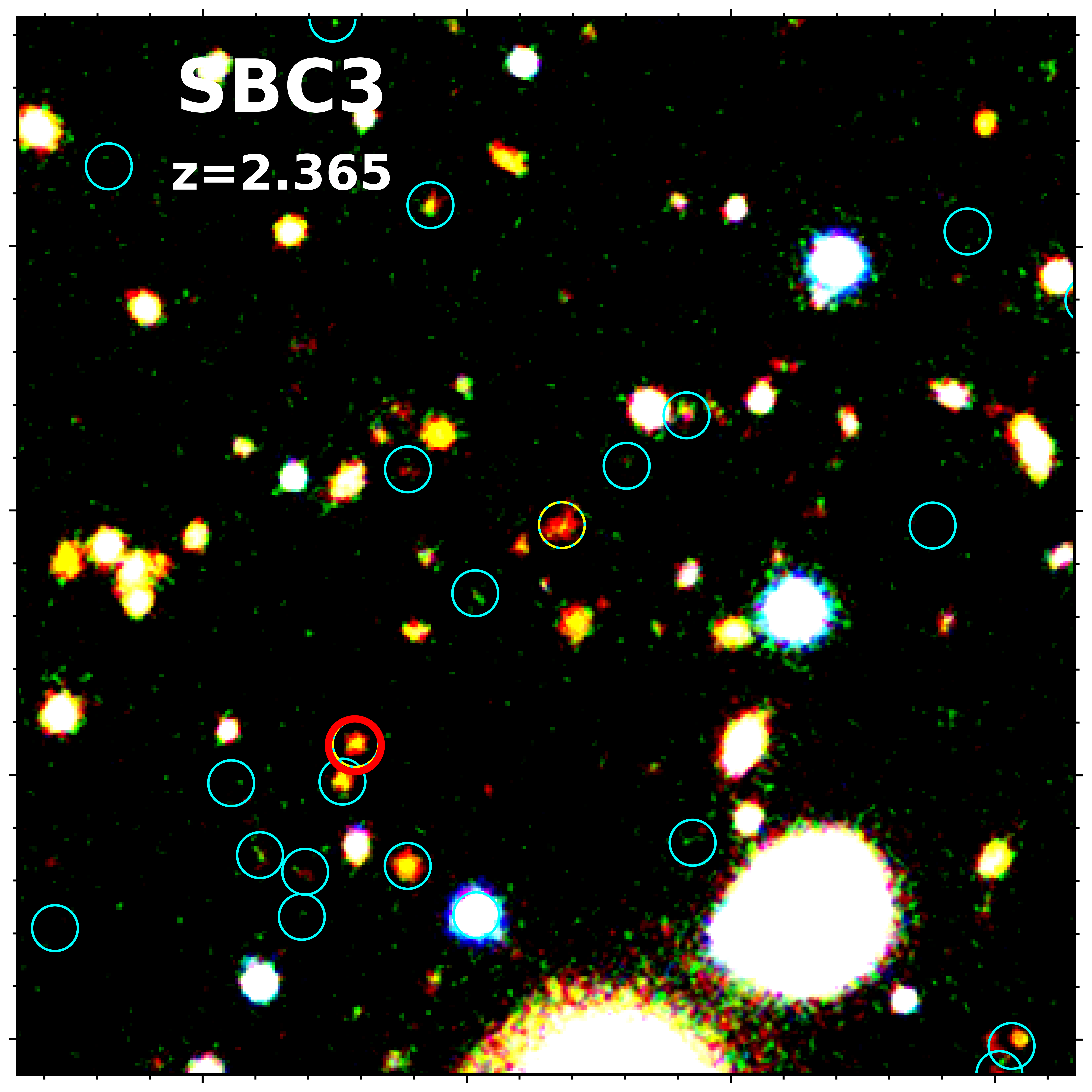}\\
\includegraphics[width=3.6in]{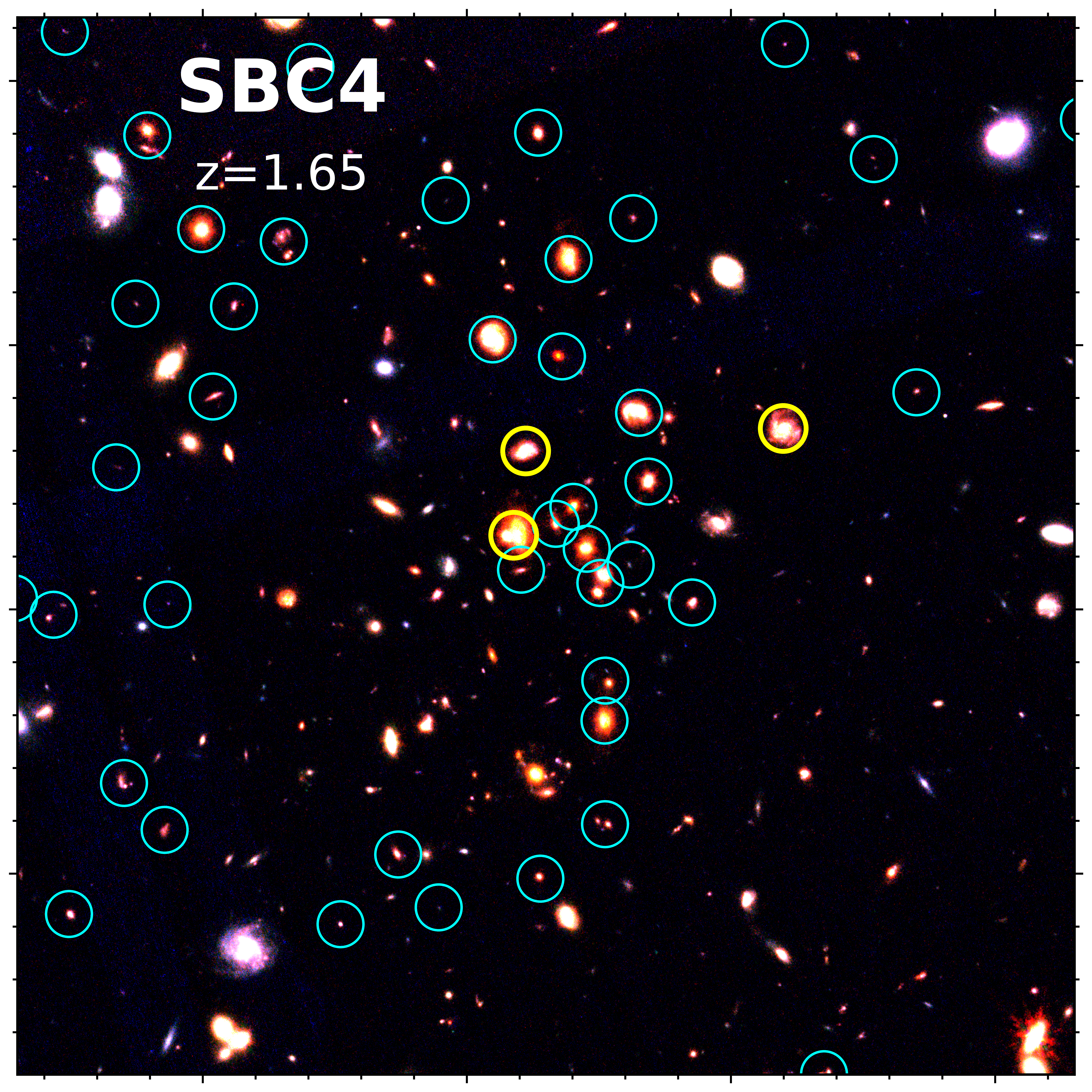} 
\includegraphics[width=3.6in]{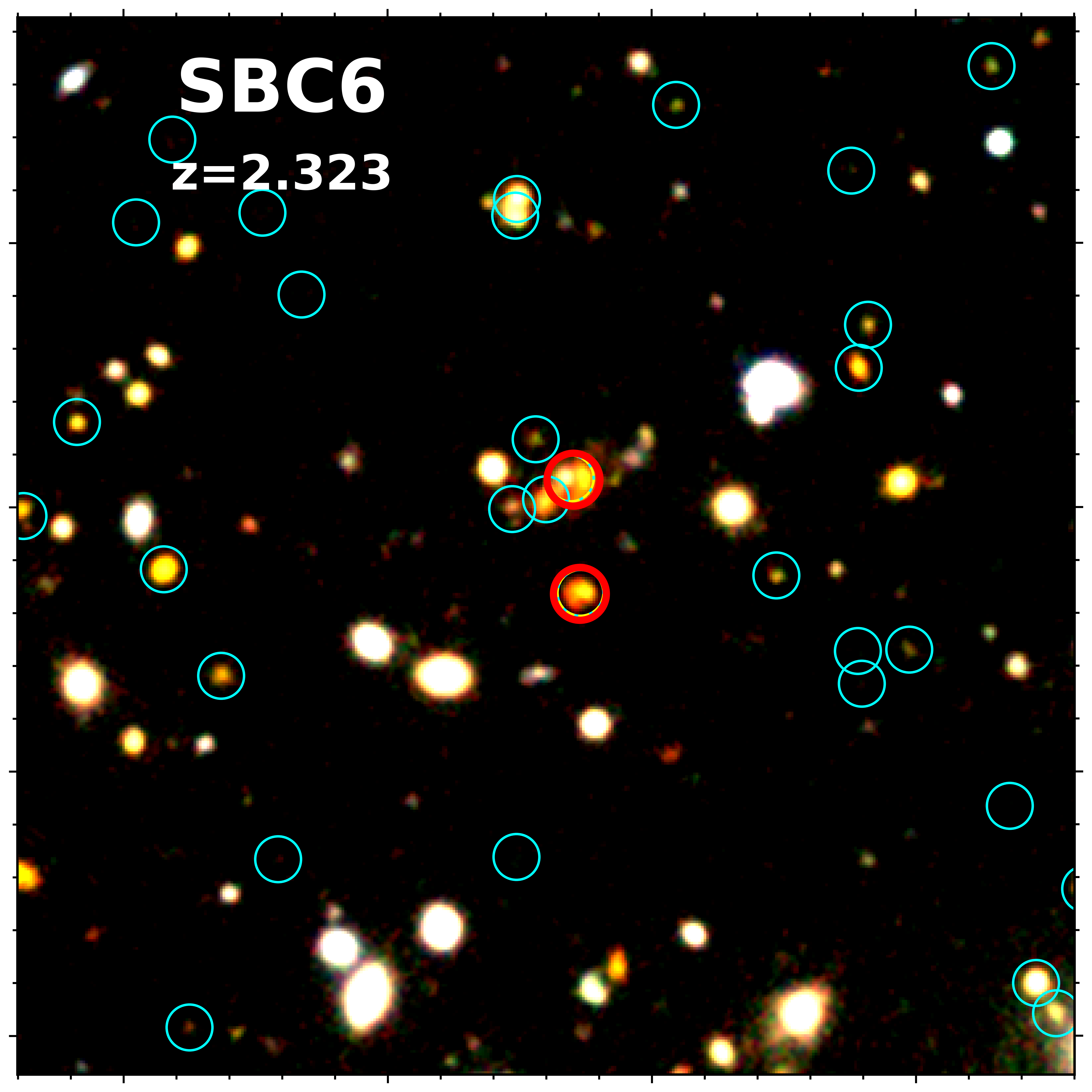}
\caption{RGB images of the eight protoclusters in the COSMOS field identified in the NICE survey. The color channels are R: [4.5], G: $K_s$, B: $Y$, except for the two protoclusters covered by COSMOS-Web \citep{Casey2023}, COSMOS-SBC4 and COSMOS-SBCX3, where the color channels are R: F444W, G: F277W, B: F150W. Each image has a size of 1\arcmin$\times$1\arcmin. The red, yellow, and cyan circles highlight the CO-detected members, spectroscopically confirmed members, and all other members identified by \citet{Sillassen2024}, respectively.}
\label{fig:img-rgb}
\end{figure}

\begin{figure}
\ContinuedFloat
\centering
\includegraphics[width=3.6in]{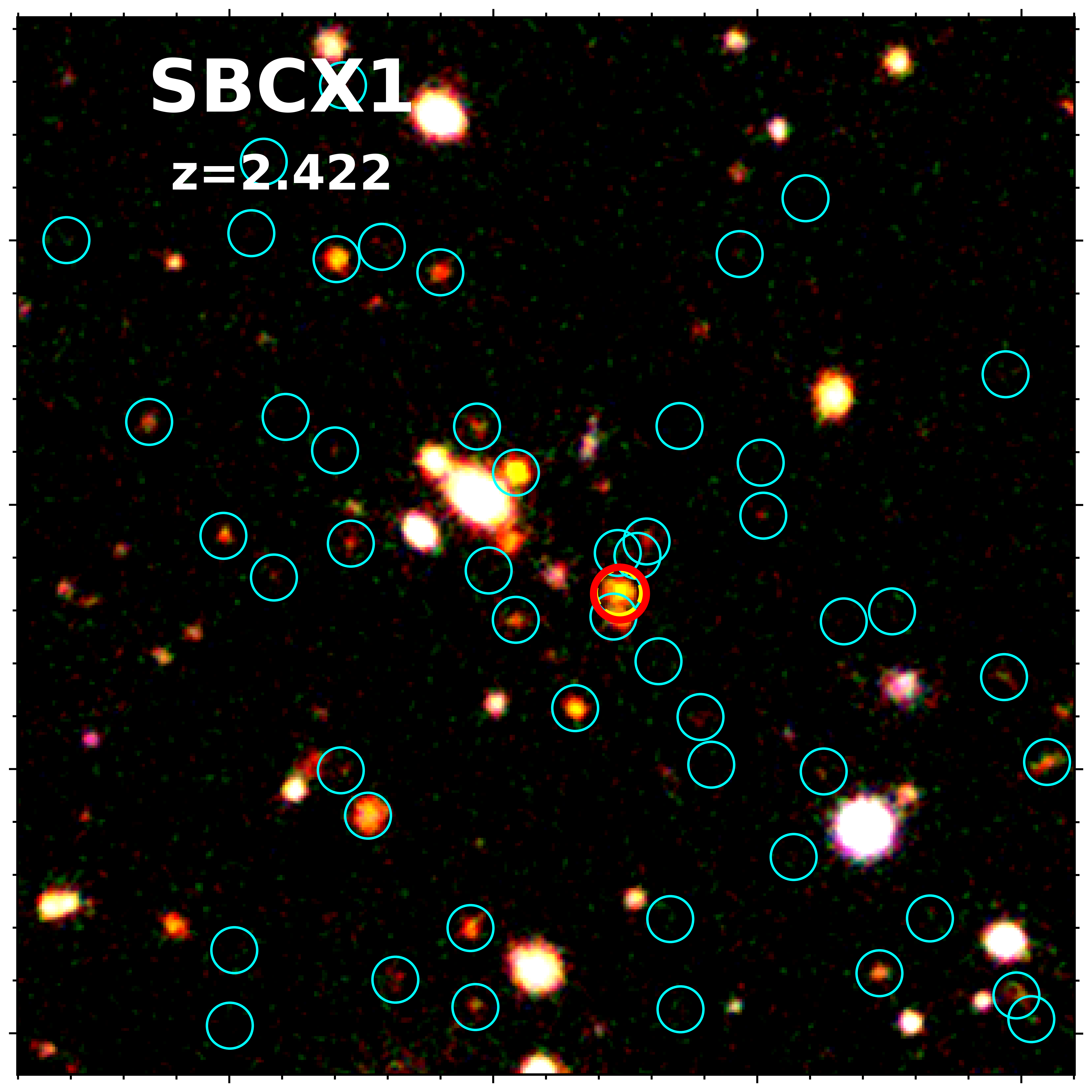}
\includegraphics[width=3.6in]{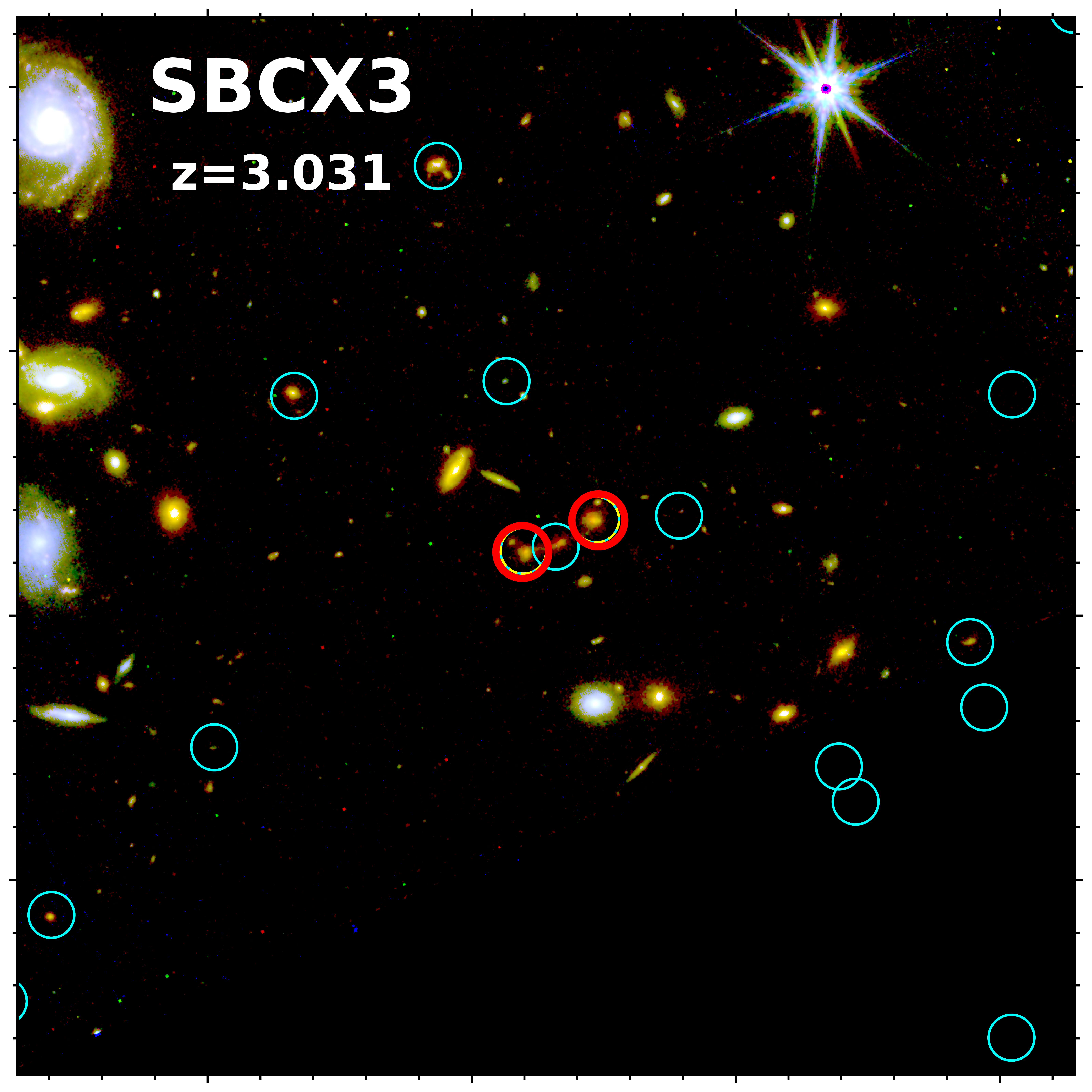} \\
\includegraphics[width=3.6in]{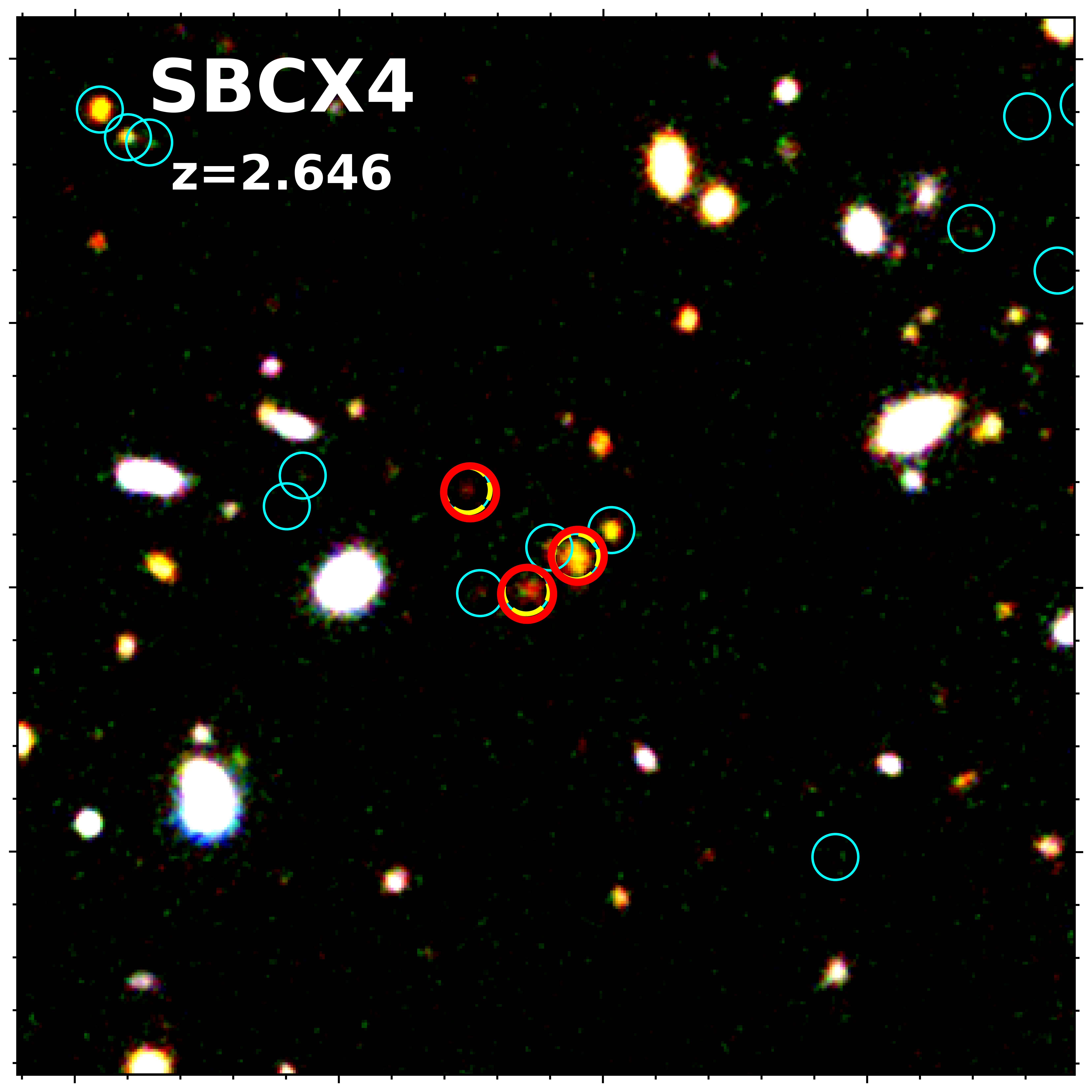}
\includegraphics[width=3.6in]{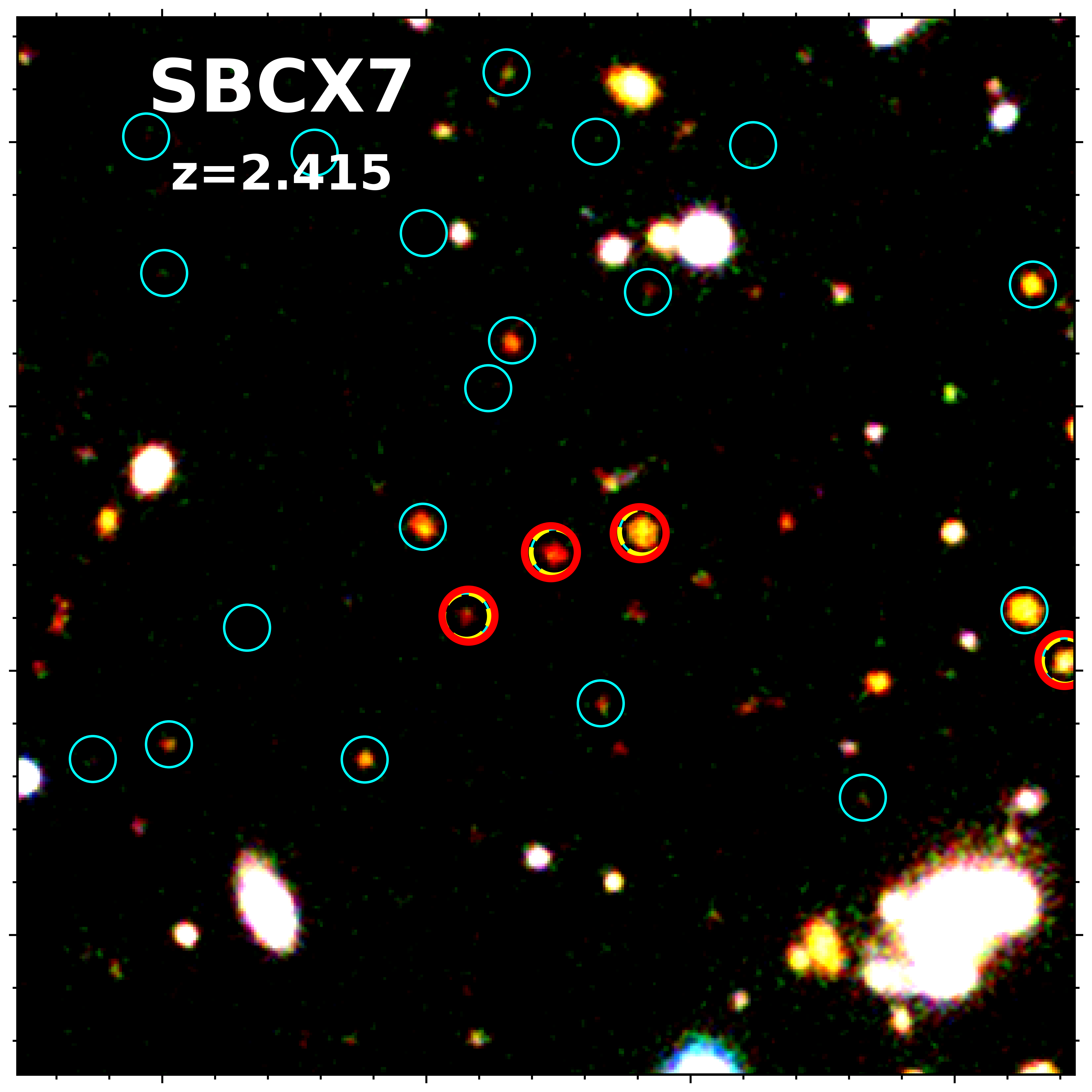}
\caption{continued.}
\label{fig:img-rgb-cont}
\end{figure}

\end{document}